\documentclass{aa}  
\usepackage{graphicx}
\usepackage{txfonts}
\usepackage{hyperref}
\usepackage{lscape}
\usepackage{threeparttable}
\usepackage{xcolor}
\usepackage{hyperref}
\usepackage[capitalise]{cleveref}

\begin{document} 

\title{JWST photometry and astrometry of 47\,Tucan\ae\ }
\subtitle{Discontinuity in the stellar sequence at the star/brown dwarf transition}

\titlerunning{47\,Tuc - JWST}
\authorrunning{M.\,Scalco et al.}

\author{M.\,Scalco\inst{1,2}\fnmsep\thanks{\email{michele.scalco@inaf.it}} 
\and
R.\,Gerasimov\inst{3}
\and
L.\,R.\,Bedin\inst{2}
\and 
E.\,Vesperini\inst{4}
\and 
M.\,Correnti\inst{5,6}
\and 
D.\,Nardiello\inst{7,2}
\and
A.\,Burgasser\inst{8}
\and
\\
H.\,Richer\inst{9}\thanks{Deceased on 13 November 2023.}
\and
I.\,Caiazzo\inst{10,11}
\and
J.\,Heyl\inst{9}
\and
M.\,Libralato\inst{2}
\and
J.\,Anderson\inst{12}
\and
M.\,Griggio\inst{12}
}

\institute{
Dipartimento di Fisica e Scienze della Terra, Università di Ferrara, Via Giuseppe Saragat 1, Ferrara I-44122, Italy
\and
Istituto Nazionale di Astrofisica, Osservatorio Astronomico di Padova, Vicolo dell’Osservatorio 5, Padova I-35122, Italy
\and
Department of Physics and Astronomy, University of Notre Dame, Nieuwland Science Hall, Notre Dame, Indiana 46556, USA
\and
Department of Astronomy, Indiana University, Swain West, 727 E. 3rd Street, Bloomington, IN 47405, USA
\and
Istituto Nazionale di Astrofisica, Osservatorio Astronomico di Roma, Via Frascati 33, I-00078, Monteporzio Catone, Rome, Italy
\and 
Agenzia Spaziale Italiana - Space Science Data Center, Via del Politecnico, I-00133, Rome, Italy
\and
Dipartimento di Fisica e Astronomia "Galileo Galilei", Universit{\`a} di Padova, Vicolo dell'Osservatorio 3, Padova I-35122, Italy
\and
Department of Astronomy \& Astrophysics, University of California, San Diego, La Jolla, California 92093, USA
\and
Department of Physics and Astronomy, University of British Columbia, 6224 Agricultural Road, Vancouver, BC V6T 1Z1, Canada
\and
Institute of Science and Technology Austria, Am Campus 1, 3400 Klosterneuburg, Austria
\and
Division of Physics, Mathematics and Astronomy, California Institute of Technology, Pasadena, CA91125, USA
\and
Space Telescope Science Institute, 3700 San Martin Drive, Baltimore, MD 21218, USA
}

\date{XXX,YYY,ZZZ}
 
\abstract
{Using JWST Near Infrared Camera (NIRCam) images of the globular cluster 47\,Tucan\ae\, (or NGC\,104), taken at two epochs just 7 months apart, we derived proper-motion membership down to $m_{\rm F322W2} \sim 27$. We identified an intriguing feature at the very low-mass end of the main sequence, around $\sim$ 0.08\,M$_{\cdot}$, at magnitudes $m_{\rm F322W2} \sim 24$ and $m_{\rm F150W2} \sim 25$. This feature, dubbed \textit{``kink''}, is characterized by a prominent discontinuity in the slope of the main sequence. A similar discontinuity is seen in theoretical isochrones with oxygen-poor chemistries, related to the rapid onset of $\mathrm{CH_4}$ absorption. We therefore hypothesize that the cluster hosts disproportionately more oxygen-poor stars near the bottom of the main sequence compared to the upper main sequence and the red giant branch. Our results show no strong or conclusive evidence of a rise in the brown dwarf luminosity function at faint magnitudes, in contrast to previous findings likely affected by faint red background galaxies. In our analysis, we accounted for this contamination by using proper motion membership.}


\maketitle

\section{Introduction}\label{Section1}
Globular clusters (GCs) are fundamental benchmarks for studying stellar evolution, as they consist of stars with similar ages, metallicities, and distances but span a wide range of masses. These systems are among the oldest known objects in the Universe and play a key role in understanding the formation and evolution of galaxies including our own, the Milky Way. GCs have been extensively studied through various observational techniques, especially with the help of space-based telescopes such as the \textit{Hubble} Space Telescope (HST), which thanks to high-resolving power, were able to observe their dense cores. HST has revolutionized our understanding of these dense stellar environments \citep{2007AJ....133.1658S} by providing unprecedented detail down to the faintest stellar members \citep[e.g.][]{2001ApJ...560L..75B}.

One of the most exciting cutting-edge areas in the study of GCs is the identification of brown dwarfs (BDs). These objects, with masses below the threshold for hydrogen fusion, evolve by cooling and dimming over time \citep{1962iss..rept....1K,1963PThPh..30..460H,1963ApJ...137.1121K}. Although thousands of BDs have been identified in young stellar clusters and the solar neighbourhood \citep[see e.g.][]{1995Natur.378..463N,1995Natur.377..129R,2010A&A...510A..27B,2012ApJ...753..156K,2019ApJS..240...19K,2021ApJ...915L...6K,2021AJ....161...42B}, only a handful have been detected in GCs \citep{2024ApJ...971...65G}, despite extensive studies with the best facilities, including HST \citep[see e.g.][]{2016ApJ...817...48D,2019MNRAS.486.2254D}. Building a larger sample of BDs in GCs would be a significant breakthrough, as it would offer insight into both the properties of BDs in different environments and the evolutionary processes at play in these ancient systems. Due to the extreme ages of GCs, these fading objects are expected to show a significant luminosity gap below the hydrogen-burning limit \citep{2004ApJS..155..191B,2017arXiv170200091C,2019BAAS...51c.521C,roman_47Tuc} which has so far remained hard to identify in GCs due to the small number of confirmed BD members.

With the advent of JWST, new opportunities have emerged for detecting BDs in GCs. JWST's infrared capabilities are prepared to probe the lowest red mass objects at the very end of the MS and into the BD regime. Recently, JWST successfully identified proper-motion BD candidates for the first time in NGC\,6397, the second closest GC \citep{2024ApJ...971...65G}.

The GC 47\,Tucan\ae \,(or NGC\,104, hereafter 47\,Tuc), with an age of $\sim$11.5\,Gyr \citep{roman_47Tuc}, is one of the closest \citep[d$\sim$4.45\,kpc;][]{dm} and most massive \citep[8.53$\times$10$^5$\,M$_{\odot}$;][]{2019MNRAS.482.5138B} clusters in the Milky Way, being relatively metal-rich \citep[\textnormal{[Fe/H]} $\sim -$0.75;][]{nominal_C14,nominal_T14,nominal_M16}. This cluster has been the subject of numerous HST studies spanning a range of wavelengths, aimed at investigating its stellar populations and internal dynamics \citep[e.g.][]{2009ApJ...697L..58A,2013ApJ...771L..15R}. JWST observations finally give us the opportunity to explore its faintest members, including potential BDs, opening a new chapter in the study of this GC.

Recent studies utilizing data from JWST have significantly enhanced our understanding of 47\,Tuc. \citet{2023MNRAS.521L..39N} have reported the first tentative discovery of BDs in 47\,Tuc, while \citet{2024ApJ...965..189M} have emphasized JWST's effectiveness in detecting multiple stellar populations (mPOPs) within the low-mass regime of M dwarfs. The findings by \citet{2024ApJ...965..189M} have revealed intriguing substructures within the lower part of the colour-magnitude diagrams (CMDs) of 47\,Tuc, including gaps and breaks that may indicate the presence of BDs, notably pointing out a luminosity function (LF) minimum associated with the hydrogen-burning limit. 

Previous JWST studies of 47\,Tuc relied on proper motion (PM) measurements derived from a combination of JWST and HST archival data. The lower sensitivity and greater positional errors of HST led to incompleteness in PM determination, particularly at faint and red magnitudes.

In this study, we leverage PMs derived ---for the first time--- from two separate JWST epochs to achieve more precise motion measurements at faint magnitudes. This enables a deeper investigation of the low-mass regime in 47\,Tuc, extending it into the BD range, and it represents a significant step forward from previous work. This effort is part of a systematic, independent reduction of archival and proprietary JWST data on 47\,Tuc, aimed at fine-tuning data reduction procedures and optimizing the positioning of fields for second-epoch observations under JWST program GO-5896 \citep{2024jwst.prop.5896S}, titled \textit{47Tuc: A Second Epoch to Extend the Mass-Rotation Relation to the Brown Dwarf Regime} \footnote{\href{https://www.stsci.edu/jwst/science-execution/program-information?id=5896}{https://www.stsci.edu/jwst/science-execution/program-information?id=5896}}. 

The paper is organized as follows: Section\,\ref{Section2} describes the data and reduction methods, while Section\,\ref{Section3} focuses on the evaluation of PMs. Section\,\ref{Section4} explains the artificial star tests (ASTs), and in Section\,\ref{Section5} we compare our results with theoretical models, offering an interpretation of the features at the lower end of the CMD of 47\,Tuc. Section\,\ref{Section6} discusses the LF, followed by a brief summary in Section\,\ref{Section7}.

\section{Data set and data reduction}\label{Section2}
The proprietary data used in this study comes from the JWST program GO-2559 \citep{2021jwst.prop.2559C}, which aims to investigate the cooling sequence (CS) of BDs and search for ancient planetary systems around white dwarfs (WDs) in 47\,Tuc. 47\,Tuc was selected for this investigation due to its favourable combination of proximity and stellar richness. While several ($\sim8$) GCs are closer to the Sun, 47\,Tuc stands out as an ideal target for studying BDs. The field chosen for this study lies in the outskirts of the cluster, where the relaxation time exceeds the age of the Universe. This ensures that low-mass stars remain largely unaffected by dynamical evolution, providing an excellent opportunity to investigate the least massive populations, including BDs. Given the fixed field of view of JWST, the massive size of 47\,Tuc provides a larger statistical sample even in its outskirts compared to the outer regions of other, closer but less populous GCs. Moreover, its proximity allows for the detection of BDs with JWST in just a few hours of observations.

\begin{centering} 
\begin{figure}
 \includegraphics[width=\columnwidth]{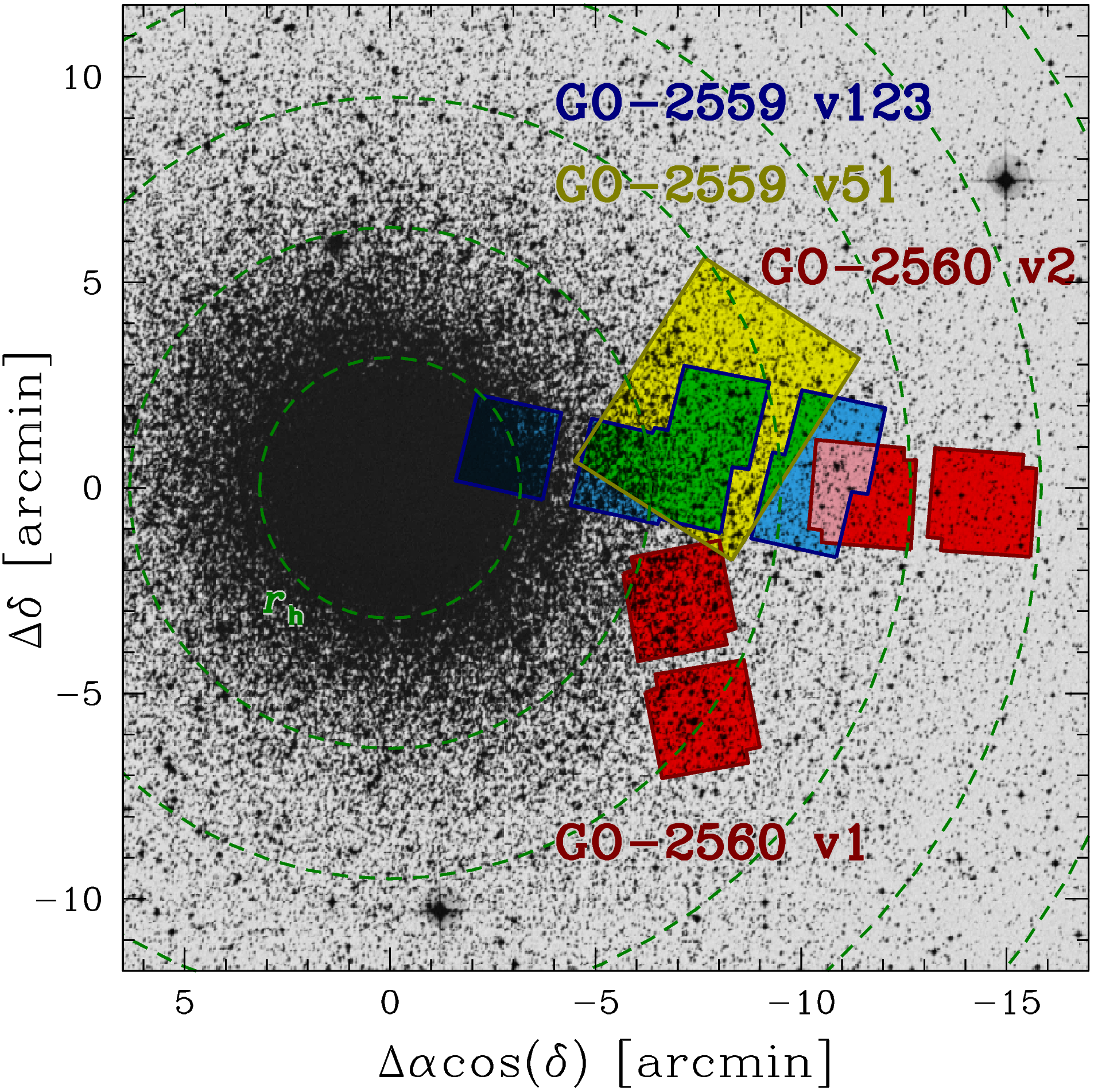}
 \caption{Outlines of the NIRCam JWST field observed under the GO-2559 programme superimposed on a DSS image of 47\,Tuc. The region observed under the first epoch (GO-2559 v123) is represented in blue, while the region observed under the second epoch (GO-2559 v51) is represented in yellow. The region of the overlap between the two epochs is highlighted in green. We also show (in red) the NIRCam JWST fields observed under the JWST GO-2560 programme \citep{2021jwst.prop.2560M}: Visit v1 \citep[see e.g.][]{2023MNRAS.521L..39N,2023MNRAS.522.2429M}, and v2 \citep[see e.g.][]{2024ApJ...965..189M}. The overlap between GO-2559 v123 and GO-2560 v2 is highlighted in magenta. The units are in arcminutes measured from the centre of the cluster. The green dashed circles mark the half-light radius ($r_{\rm h}=3^\prime_{\cdot}17$; \citealt{1996AJ....112.1487H,2010arXiv1012.3224H}) with additional circles marking $n\,r_{\rm h}$, where $n=2,...,6$.}
 \label{FOV} 
\end{figure} 
\end{centering}

The data for this article were collected using the Near Infrared Camera \citep[NIRCam,][]{2023PASP..135b8001R} on JWST, focusing on a region located at an average angular distance of $\sim$7\,arcmin from the centre of 47\,Tuc. Figure \ref{FOV} shows the position of the field-of-view (FOV) superimposed on a Digital Sky Survey\footnote{\href{https://archive.eso.org/dss/dss}{https://archive.eso.org/dss/dss}} (DSS) image of 47\,Tuc. 

The observations were carried out simultaneously using the Short Wavelength (SW) and Long Wavelength (LW) channels over two epochs: September 14-15, 2022 (epoch GO-2559 v123, $\sim$2022.7) and May 24-25, 2023 (epoch GO-2559 v51, $\sim$2023.4). Ultra-wide filters were applied in both NIRCam channels, with F150W2 for the SW channel and F322W2 for the LW channel.

During the first epoch, three visits were conducted without a primary dither, but with a \texttt{STANDARD} sub-pixel dither pattern of 16 sub-pixel positions. At each pointing, a single image was taken in both channels using the \texttt{MEDIUM8} readout pattern (8 groups), resulting in a total of 96 images with the F322W2 filter and 384 images with the F150W2 filter, with an effective exposure time of 837.468\,s per image. A fourth visit (v4) failed due to the mis-acquisition of the guide stars, caused by the incomplete state of the guide star catalogue during the early operational phase of JWST. 

This failed visit (v4) was re-acquired approximately 7 months later and designated as visit 51 (v51). This second epoch is a single visit and was conducted using a 6-point \texttt{FULLBOX} primary pattern to cover the gaps between detectors within a module, combined with a \texttt{STANDARD} sub-pixel dither pattern with 1 sub-pixel position. At each pointing, a single image was captured in both channels using the \texttt{MEDIUM8} readout pattern (8 groups), producing 24 images with the F322W2 filter and 96 images with the F150W2 filter, with an effective exposure time of 944.836\,s per image.

The image reduction followed the procedures outlined in similar studies \citep[see e.g.,][]{2024AN....34540039B, 2024A&A...689A..59S,2022MNRAS.517..484N, 2023MNRAS.525.2585N, 2023MNRAS.521L..39N, 2023AN....34430006G, 2023ApJ...950..101L, 2024PASP..136c4502L}. We first processed the level-1b uncalibrated (\texttt{\_uncal}) images using a development version of the JWST pipeline\footnote{\href{https://github.com/spacetelescope/jwst}{https://github.com/spacetelescope/jwst}} \citep{2023BushouseJWSTpipeline} through stages 1 and 2 to obtain the level-2b (\texttt{\_cal}) images.

The reduction of the \texttt{\_cal} images consists of a first-pass and second-pass photometry. The first-pass photometry collects PSFs, positions, and magnitudes of the stars in each single image. We applied our current best geometric distortion correction to the star positions \citep{2023AN....34430006G}. The star coordinates from each image were then transformed into a common reference frame, using {\it Gaia} Data Release 3 \citep[DR3, ][]{2016A&A...595A...1G,2023A&A...674A...1G} bright members as the reference.

In the second-pass photometry, we used a modified version of the \texttt{KS2} code developed by J.\ Anderson and described in many previous studies 
\citep[][and references therein]{2017ApJ...842....6B,2018ApJ...853...86B,2018MNRAS.481.3382N,2018ApJ...854...45L,2022LibralatoPMcat,2021MNRAS.505.3549S}. In this step, we extracted positions and fluxes using both the PSFs and the transformations derived in the first-pass photometry. \texttt{KS2} utilizes all images simultaneously, making it particularly suitable for obtaining deep photometry of sources too faint to be detected in individual images. Along with fluxes and positions, \texttt{KS2} produces several quality diagnostics, including the PSF quality of fit (QFIT) parameter, the RADXS parameter \citep[a parameter that describes how well a given source’s shape resembles the PSF, making it useful for distinguishing between galaxies and star-like sources; see][]{2008ApJ...678.1279B,2009ApJ...697..965B}, and the local sky noise (rmsSKY). For a detailed description of these parameters, we refer to \citet{2018ApJ...853...86B,2021MNRAS.505.3549S,2018MNRAS.481.3382N}.

We calibrated the photometry to the VEGA-magnitude system following the procedures illustrated in \citet{2005MNRAS.357.1038B} and anchored the astrometry to the International Celestial Reference System (ICRS) frame using \textit{Gaia} DR3 data for sources in the observed field.

We corrected our photometry for the effects of differential reddening and photometric zero-point variations across the FOV, following the method and the standard procedures adopted in many previous studies \citep[e.g.,][]{2007AJ....133.1658S,2012A&A...540A..16M,2017ApJ...842....7B}.

We reduced the data from the two epochs separately. Since the first epoch provides deeper and higher-quality data, we will use the photometry from the first epoch moving forward. The data from the second epoch was used to register the photometry of the many detectors into a common reference system (as v51 links stars measured in different detectors to the same detector) and it will be used exclusively to evaluate PMs.

\begin{centering} 
\begin{figure*}
 \includegraphics[height=0.49\textwidth]{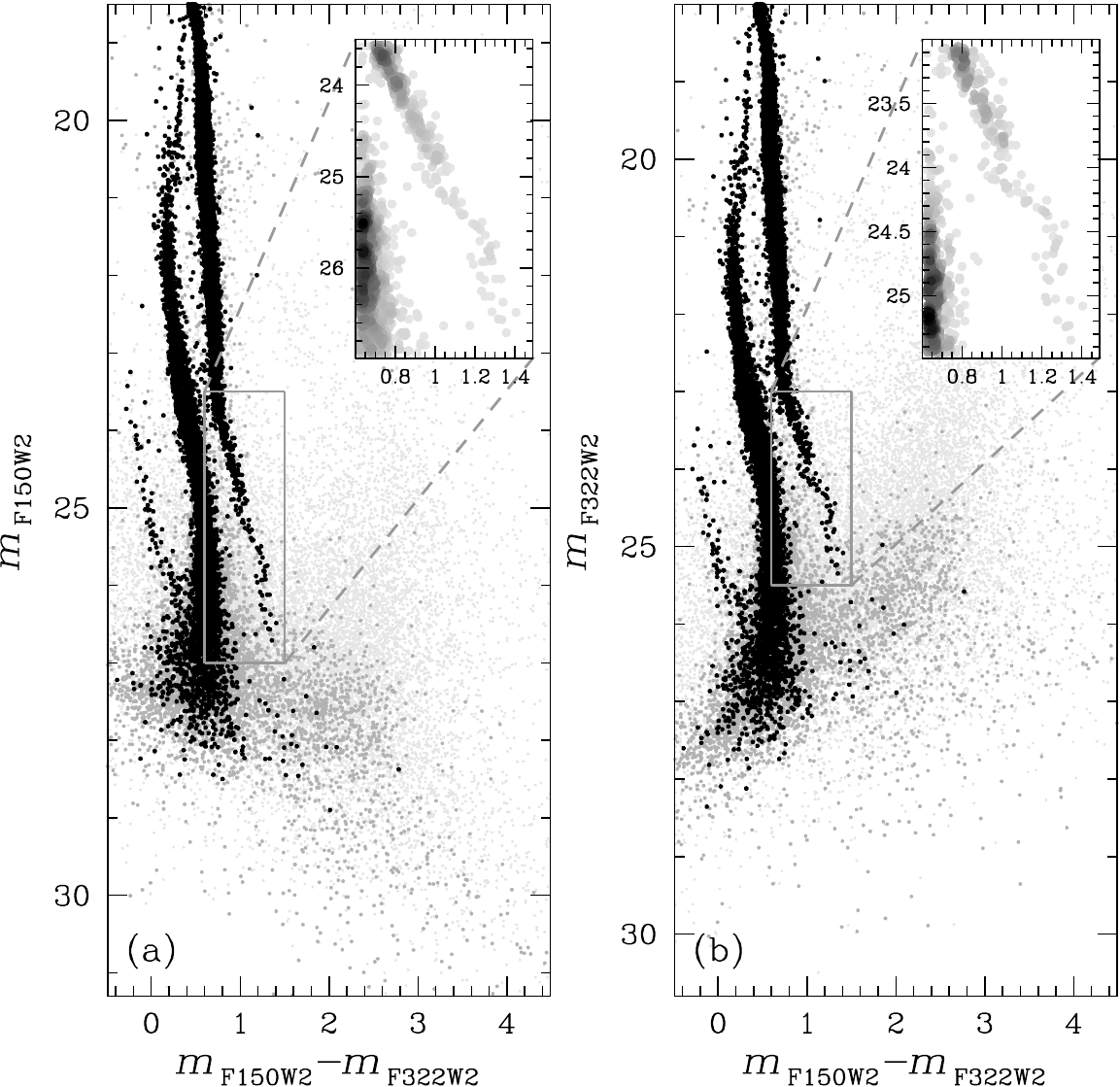}
 \includegraphics[height=0.49\textwidth]{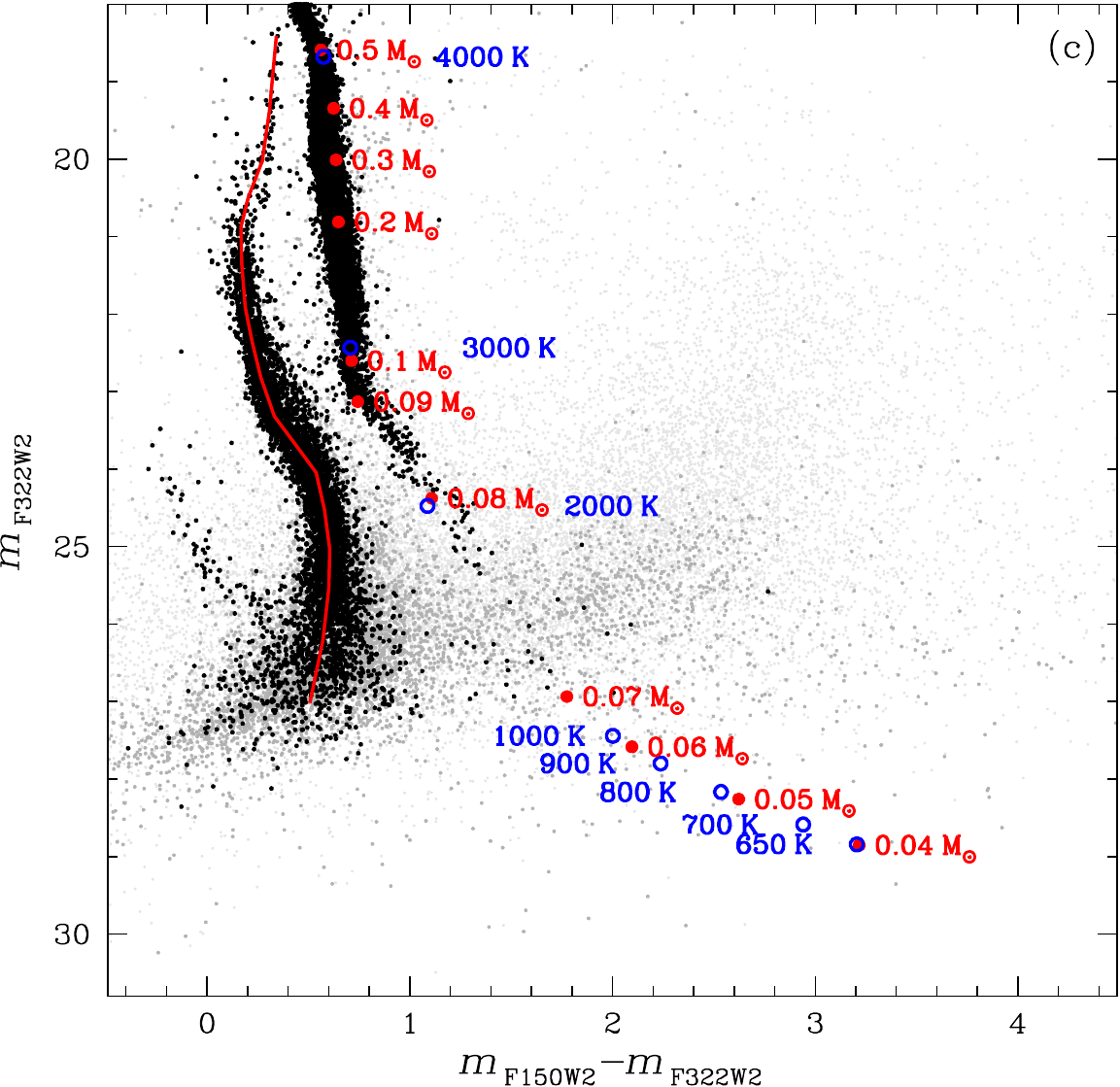}
 \caption{Colour-magnitude diagrams of 47\,Tuc and the SMC using JWST filters. (a) $m_{\rm F150W2}$ versus $m_{\rm F150W2}-m_{\rm F322W2}$ CMD. (b) $m_{\rm F322W2}$ versus $m_{\rm F150W2}-m_{\rm F322W2}$ CMD. (c) Same as (b), but with overlaid the expected mass values (in red) and expected $T_{\rm eff}$ values (in blue), based on model isochrones from \citet{roman_47Tuc} and discussed in Section\,\ref{Section5}, specifically the \texttt{G24R} model isochrones (see Section\,\ref{Section5} for further details). The red fiducial line was manually defined to mark the position of the SMC MS on the CMD. In all panels, black dots represent stars that satisfy all selection criteria, while light grey dots indicate stars that do not meet the QFIT or rmsSKY criteria. Dark grey dots denote stars that pass the QFIT and rmsSKY criteria but do not satisfy the RADXS selection.} 
 \label{CMD} 
\end{figure*} 
\end{centering}

We selected a sample of well-measured stars using the quality parameters provided by \texttt{KS2} (QFIT, RADXS, and rmsSKY), following a similar approach to that described in \citet{2021MNRAS.505.3549S,2024AN....34540018S,2018MNRAS.481.3382N,2017ApJ...842....6B}. Figure\,\ref{CMD} presents the $m_{\rm F150W2}$ versus $m_{\rm F150W2}-m_{\rm F322W2}$ colour-magnitude diagram (CMD, panel a) and the $m_{\rm F322W2}$ versus $m_{\rm F150W2}-m_{\rm F322W2}$ CMD (panel b) for all stars in the field of the first epoch. Following the referee's suggestion, panel\,(c) of Fig.\,\ref{CMD} presents the $m_{\rm F322W2}$ versus $m_{\rm F150W2} - m_{\rm F322W2}$ CMD (shown in panel b), with overlaid mass and effective temperature ($T_{\rm eff}$) values derived from the model isochrones presented in \citet{roman_47Tuc} and discussed in Section\,\ref{Section5}, specifically the \texttt{G24R} model isochrones (see Section\,\ref{Section5} for further discussion). In all panels, black dots represent stars that meet all selection criteria, while light grey dots indicate stars that fail to meet the QFIT or rmsSKY criteria. Dark grey dots denote stars that pass the QFIT and rmsSKY criteria but fail the RADXS selection. Notably, the dark grey points form a visible clump of sources at $m_{\rm F150W2}\sim27.5$ ($m_{\rm F322W2}\sim25.5$) and $m_{\rm F150W2}-m_{\rm F322W2}\sim1.8$, likely corresponding to galaxies and artefacts, which are effectively filtered out by the RADXS parameter.

Focusing on the black points, we observe the clear emergence of the cluster’s lower main sequence (MS), the WD sequence, and the Small Magellanic Cloud (SMC) MS. To indicate the position of the SMC MS in the plot, we manually defined its fiducial line, shown in red in panel (c). One of the most striking aspects of these CMDs is the intricate and detailed structure of the lower portion of the MS, which was presented by \citet{2024ApJ...965..189M}. This complex structure is particularly emphasized in the two insets, which provide a zoomed-in view of the Hess diagram focusing on this region of the lower MS. 

Several notable features, previously discussed by \citet{2024ApJ...965..189M}, are evident in the overall CMDs: a clump around $m_{\rm F322W2} \sim 23.7$, a gap near $m_{\rm F322W2} \sim 24$, and a change in slope at $m_{\rm F322W2} \sim 24.5$. The sequence appears to terminate around $m_{\rm F322W2} \sim 25.5$, followed by a sparse group of sources extending down to $m_{\rm F322W2} \sim 27$. While the sources above $m_{\rm F322W2} \sim 25.5$ are clearly stars, the sources below $m_{\rm F322W2} \sim 25.5$ may represent the bright part of the BDs CS, as suggested by \citet{2024ApJ...965..189M}. However, it is important to note that they fall within a region of the CMD typically populated by galaxies. While most of these resolved objects were effectively filtered out using the RADXS parameter (see dark grey points in Fig.\,\ref{CMD}), the remaining sources may include point-like galaxies that escaped this selection. This underscores the need for precise PMs at these faint magnitudes to accurately establish the cluster membership of these sources.

In the next section, we will undertake a detailed analysis of the PMs for these stars, utilizing data from two JWST observational epochs. By evaluating their PMs, we aim to determine the cluster membership of stars in this lower portion of the CMD and provide further insight into their nature.

\section{Proper motions}\label{Section3}
\begin{centering} 
\begin{figure*}
 \includegraphics[width=\textwidth]{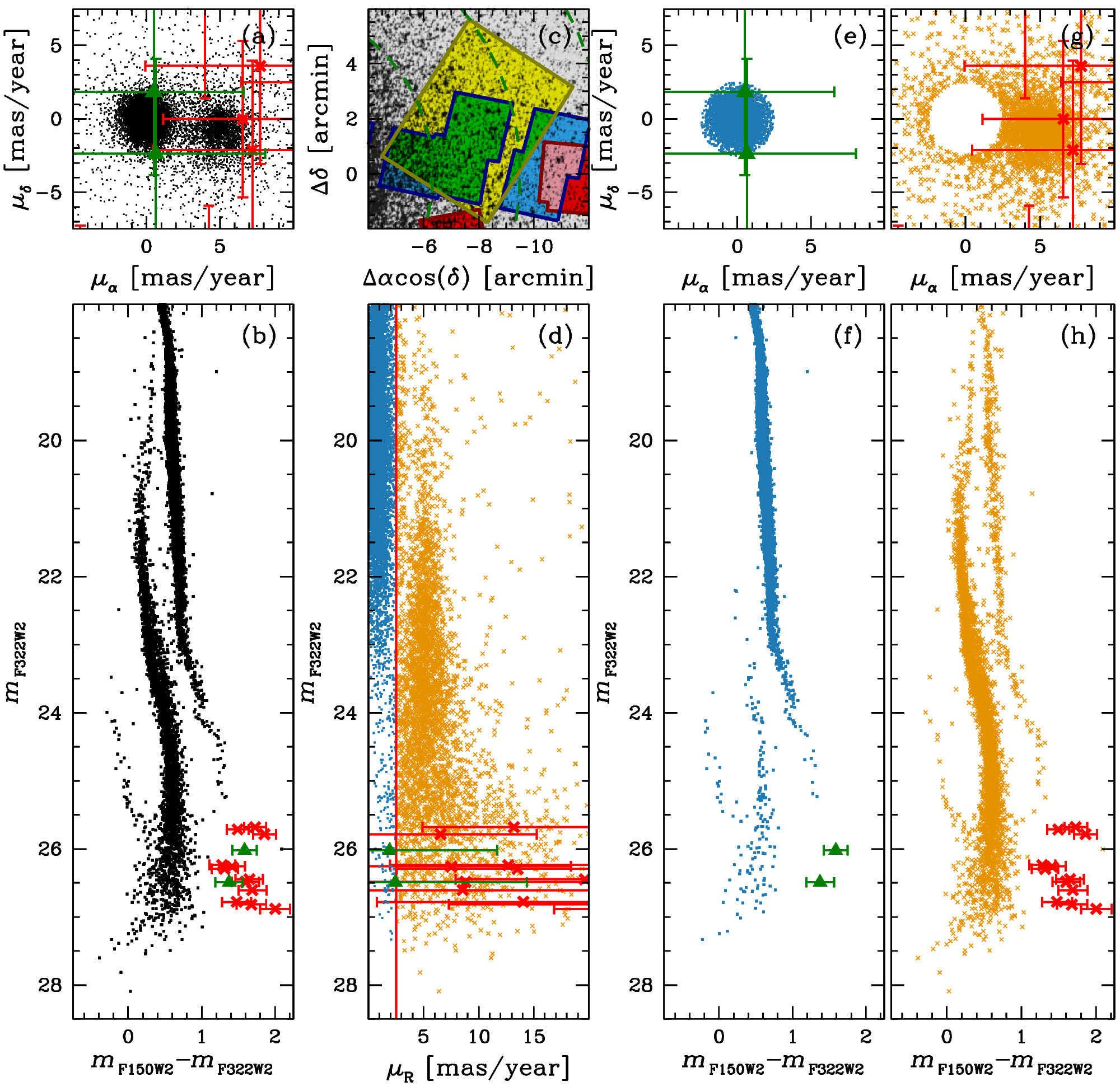}
 \caption{Cluster membership selection based on PMs obtained from displacements from JWST observations from two epochs collected under program GO-2559 over a period of just 7\,months. (a)-(b) VPD and $m_{\rm F150W2}$ versus $m_{\rm F150W2}-m_{\rm F322W2}$ CMD, respectively, for sources with measurable PMs. (c) FOV zoomed on the region of the overlap between the two epochs, represented in green. (d) $m_{\rm F150W2}$ magnitude versus the one-dimensional PM ($\mu_{\rm R}$). The red line separates cluster members from SMC and field stars. (e)-(f) VPD and $m_{\rm F150W2}$ versus $m_{\rm F150W2}-m_{\rm F322W2}$ CMD for stars that passed the PM selection. (g)-(h) same diagrams for stars that did not pass the PM selection. In panels (d) through (h), sources that passed the PM selection are represented by blue dots, while those that did not pass are depicted as orange crosses. Candidate BDs that passed and did not pass the PM selection are represented as green triangles and red crosses, respectively. The green and red error bars illustrate uncertainties: PM errors in (a), (e), and (g); colour errors in (b), (f), and (h); and errors in $\mu_R$ in (d). The uncertainties are estimated as detailed in Section\,\ref{Section4}.} 
 \label{PM} 
\end{figure*} 
\end{centering}

Proper motions were computed as displacements between the two epochs divided by the temporal baseline ($\sim$7\,months). Panels\,(a) and (b) of Fig.\,\ref{PM} show the vector-point diagram (VPD) and the $m_{\rm F322W2}$ versus $m_{\rm F150W2}-m_{\rm F322W2}$ CMD, respectively, for the sources with measurable PMs. Since PMs are calculated relative to the cluster's overall motion, the distribution of cluster members in the VPD is centred at (0,0). A second distinct cluster of points, visible around (5,-0.5), represents stars belonging to the SMC. Panel\,(c) offers a zoomed-in view of the FOV shown in Fig.\,\ref{FOV}, focused on the overlapping region between the two epochs (highlighted in green). Panel\,(d) shows the one-dimensional PM ($\mu_{\rm R}$, obtained by summing the PMs in the two directions in quadrature) plotted against $m_{\rm F322W2}$. 47\,Tuc members exhibit a narrow distribution in $\mu_{\rm R}$, mostly clustered below $\mu_{\rm R}<2.5$ mas yr$^{-1}$, while the Galactic-field objects extend towards higher $\mu_{\rm R}$, peaking around $\mu_{\rm R} \sim 5$ mas yr$^{-1}$, for the bulk of the SMC objects in the background. The distribution broadens at fainter magnitudes due to increasing random errors between the two epochs, making the separation between the cluster and the SMC less distinct. We defined a PM selection criterion to separate cluster members from SMC stars and field objects, indicated by the vertical red line at 2.5\,mas\,yr$^{-1}$. Panels\,(e) and (f) show the VPD and $m_{\rm F322W2}$ versus $m_{\rm F150W2}-m_{\rm F322W2}$ CMD for the stars that passed the PM selection, while panels\,(g) and (h) show the same diagrams for stars that did not pass the PM selection. 

From panel\,(f) we can see that most of SMC's stars are removed, but clearly, several SMC's members have PMs with uncertainties large enough to fall within the cluster-membership criterion set by the red vertical line defined in panel\,(d). Notably, the intricate structure of the lower MS remains evident after applying the PM selection. Most stars fainter than $m_{\rm F322W2} \sim 25.5$ have been filtered out; these removed sources are represented as red crosses in panels\,(a), (b), (d), (g), and (h). Only two sources (marked as green triangles in panels\,(a), (b), (d), (e), and (f)) remain. For both the red crosses (removed stars) and the green triangles (stars passing the PM selection), the uncertainties (evaluated as in Sect.\,\ref{Section4}) are shown as error bars (PM errors in panels\,(a), (e), and (g); colour errors in panels\,(b), (f), and (h); and $\mu_R$ errors in panel\,(d)).

Although the low photometric errors for the two remaining sources might suggest that they are indeed BDs, their large PM uncertainties (shown in panels\,(a), (d) and (e)) at such faint magnitudes make it challenging to definitively confirm their cluster membership. It is possible that these two objects are red galaxies, with PM errors significant enough to mistakenly place them within the cluster-membership boundary. The large uncertainties make any conclusions regarding the cluster membership of these two sources highly unreliable.

In the next section, we will make use of ASTs to examine the completeness of our star sample and assess the reliability of these two sources.
 
In Fig.\,\ref{FOV}, we can see that our JWST GO-2559 dataset partially overlaps with a field from another JWST program, GO-2560 \citep{2021jwst.prop.2560M}, particularly with the second visit of that program (labelled GO-2560 v2 in Fig.\,\ref{FOV}). The GO-2560 v2 dataset has been thoroughly studied in \citet{2024ApJ...965..189M}, where in-depth analyses have been performed. The images from the GO-2560 v2 field were taken on 24 September 2023 (epoch $\sim$2023.7), nearly a year after the images from the GO-2559 v123 dataset were captured. This temporal baseline provides an excellent opportunity to evaluate PMs in this region, allowing us to extend our study in terms of the area covered.

\begin{centering} 
\begin{figure}
 \includegraphics[width=\columnwidth]{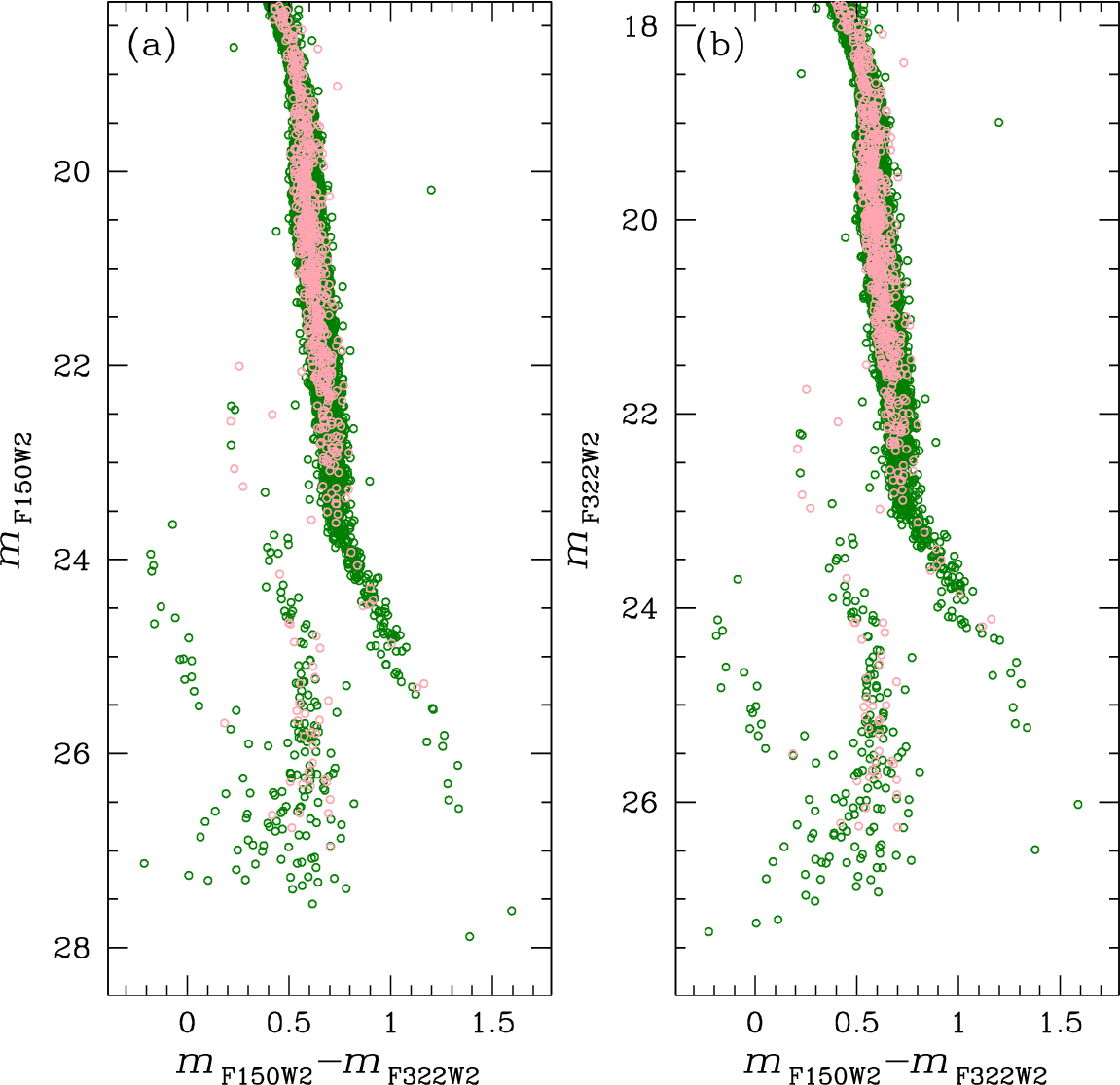}
 \caption{(a) $m_{\rm F150W2}$ versus $m_{\rm F150W2}-m_{\rm F322W2}$ CMD for the selected sample of stars, with green circles representing stars located in the region of overlap between the two GO-2559 epochs, and magenta circles indicating stars in the overlap region between GO-2559 v123 and GO-2560 v2. (b) Same as (a), but for the $m_{\rm F322W2}$ versus $m_{\rm F150W2}-m_{\rm F322W2}$ CMD.} 
 \label{comp} 
\end{figure} 
\end{centering}

We processed all the images from the GO-2560 v2 dataset following the same reduction procedures as outlined earlier. We then evaluated PMs by calculating the displacement of stars between the two epochs and dividing by the temporal baseline ($\sim$1 year). Using the same selection criteria described in Fig.\,\ref{PM}, we identified a sample of probable cluster members. Fig.\,\ref{comp} presents the $m_{\rm F150W2}$ versus $m_{\rm F150W2}-m_{\rm F322W2}$ CMD (panel (a)) and the $m_{\rm F322W2}$ versus $m_{\rm F150W2}-m_{\rm F322W2}$ CMD (panel (b)) for the selected stars in the overlapping region between GO-2559 v123 and GO-2560 v2 (represented by magenta circles). For comparison, we also show the stars from the overlap between the two GO-2559 epochs (green circles).

As shown, the number of stars represented by the magenta circles is noticeably smaller compared to the green circles, and the stars in the magenta circles appear to be less deep in terms of magnitude. The overlap region between GO-2559 v123 and GO-2560 v2 accounts for only about 20\% of the area of the overlap between the two GO-2559 epochs and is located farther from the cluster centre. As a result, this more peripheral region yielded a smaller number of detected stars. Due to the reduced number of stars and their insufficient depth, this region does not contribute significantly to our analysis. Therefore, we decided to exclude this region from the following analyses in this article and focus exclusively on the stars from the overlap between the two GO-2559 epochs.

\section{Artificial stars}\label{Section4}
We performed ASTs to assess the completeness and to estimate photometric and astrometric errors for the stars in our sample. A total of 10$^5$ artificial stars (ASs) were generated, uniformly distributed across the common FoV between the two GO-2559 epochs. We assumed a uniform distribution in F322W2 magnitudes, with the corresponding F150W2 magnitudes assigned based on a fiducial line manually defined in the $m_{\rm F150W2}$ versus $m_{\rm F150W2}-m_{\rm F322W2}$ CMD. This fiducial line traces the MS and lower MS, following the features observed in the CMDs presented in earlier sections, extending to the apparent end of the sources and extrapolating to fainter magnitudes.

Using \texttt{KS2}, we generated, detected, and measured these ASs following the same procedure as for the real stars. An AS was considered successfully recovered if the difference between its input and output positions was less than 1 pixel, the difference in magnitudes was within 0.75 (equivalent to $\sim$2.5log2) in both filters and if it passed the same selection criteria applied to real stars. To incorporate PM selection into our completeness evaluation, we ran the ASs separately for each epoch and then assessed the displacement of the stars between the two epochs. Artificial stars are generated to have identical positions in both epochs; therefore, they have no intrinsic displacement. Any observed displacement in the ASs is solely due to spurious offsets introduced by noise in the PSF shape, which can cause slight shifts in the recovered positions across epochs. For an AS to be considered recovered, it also needs to satisfy the same PM selection criteria that were applied to real stars, as shown in panel\,(d) of Fig.\,\ref{PM}. By applying this PM selection to ASs, we estimate the fraction of real stars excluded due to these positional offsets, allowing us to incorporate this effect into the overall completeness calculation.

\begin{centering} 
\begin{figure*}
 \includegraphics[width=\textwidth]{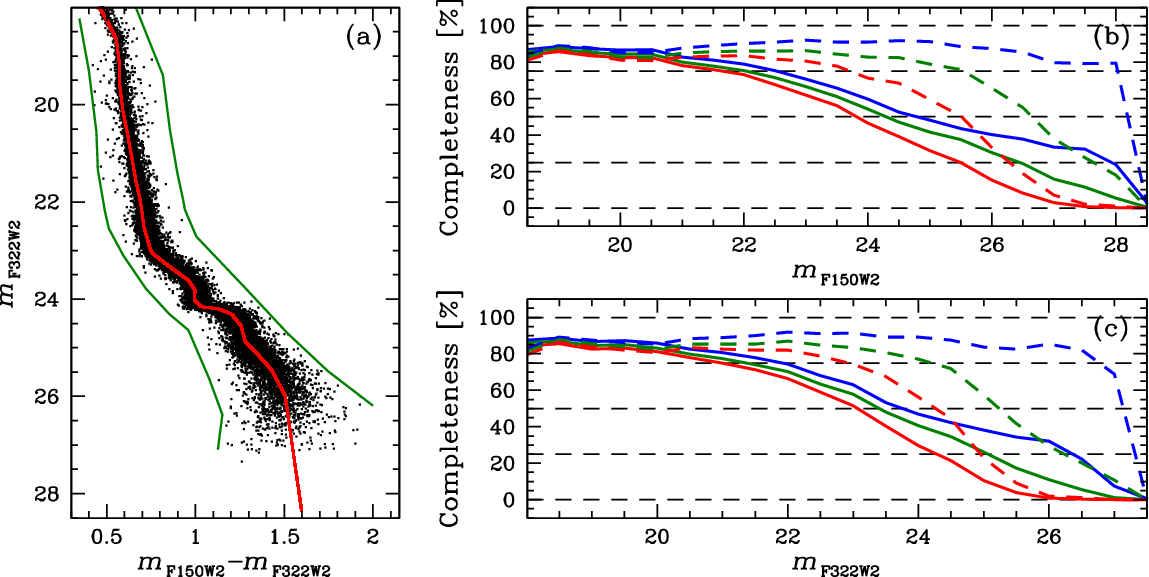}
 \caption{(a) Injected (red) and recovered (black) ASs in the $m_{\rm F322W2}$ versus $m_{\rm F150W2}-m_{\rm F322W2}$ CMD. The two green lines define the boundaries of the MS. (b)-(c) Completeness as a function of $m_{\rm F150W2}$ and $m_{\rm F322W2}$ magnitudes, respectively, with (red solid line) and without (green solid line) applying the PM selection, as well as without considering either the PM or RADXS selections (blue solid line). The dashed lines represent the completeness values in regions limited to the dark usable areas for detecting faint sources.} 
 \label{ASs} 
\end{figure*} 
\end{centering}

Panel\,(a) of Fig.\,\ref{ASs} shows the injected and recovered ASs in the $m_{\rm F322W2}$ versus $m_{\rm F150W2}-m_{\rm F322W2}$ CMD, with red points representing the injected stars and black points indicating the recovered ones. We manually defined two fiducial lines that enclose the MS and classified all stars outside this region as unrecovered. Panels\,(b) and (c) illustrate, in red, the completeness levels derived from our AST, defined as the ratio of injected to recovered stars, as a function of the $m_{\rm F150W2}$ and $m_{\rm F322W2}$ magnitudes, respectively. For comparison, we also show in green the completeness levels obtained without considering the PM selection, and in blue those derived without applying either the PM or RADXS selections. The dashed lines represent $c_g$, the completeness values in the regions constrained to the dark usable areas for detecting faint sources \citep[see][for more details]{2008ApJ...678.1279B,2009ApJ...697..965B}.

We evaluated the uncertainties in our PMs by analyzing the dispersion of the ASs displacements between the two epochs as a function of magnitude. Similarly, we estimated the photometric uncertainties by examining the dispersion between the inserted and recovered magnitudes, also as a function of magnitude. The uncertainties for the candidate BDs identified in this study are displayed in Fig.\,\ref{PM} (see Section\,\ref{Section3} for details). 
 
\begin{centering} 
\begin{figure}
 \includegraphics[width=\columnwidth]{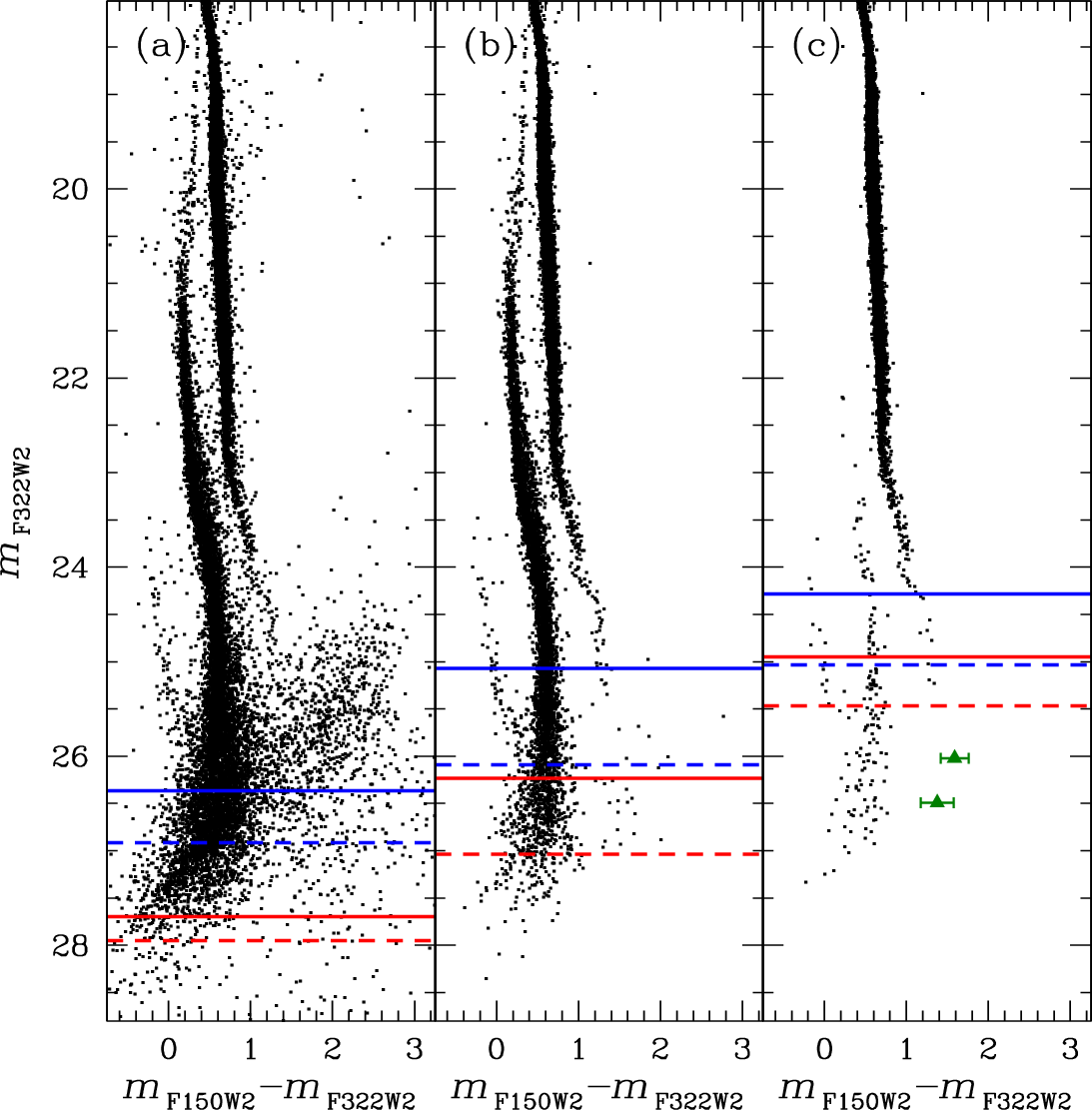}
 \caption{The $m_{\rm F322W2}$ versus $m_{\rm F150W2}-m_{\rm F322W2}$ CMDs are shown for three cases: without both PM and RADXS selections (a), with all photometric selections (b), and with both photometric and PM selections (c). The solid and dashed blue horizontal lines indicate the magnitudes at which completeness reaches 25\% and 10\%, respectively, across all CMDs, while the solid and dashed red lines represent the same for $c_g$. In (c), the two suspected BDs are highlighted as green triangles, with green error bars indicating their associated uncertainties.} 
 \label{ASs2} 
\end{figure} 
\end{centering}

Figure\,\ref{ASs2} shows the $m_{\rm F322W2}$ versus $m_{\rm F150W2}-m_{\rm F322W2}$ CMD, comparing different selection cases: panel\,(a) presents the CMD without both PM and RADXS selections, panel\,(b) applies all photometric selections, and panel\,(c) includes both photometric and PM selections. In all panels, the blue solid and dashed lines represent the magnitudes where the completeness drops to 25\% and 10\%, respectively, while the red solid and dashed lines show the same for $c_g$.

In the CMDs without PM selection (panels a and b), we observe that completeness remains relatively high at fainter magnitudes. In panel\,(b), completeness is 25\% at $m_{\rm F322W2}\sim25$ and 10\% at $m_{\rm F322W2}\sim26$. This explains why we see a significant number of sources extending down to $m_{\rm F322W2}\sim27$ at the lower end of the MS. In contrast, the CMD with PM selection (panel c) shows lower completeness, with values of 25\% at $m_{\rm F322W2}\sim24.3$ and 10\% at $m_{\rm F322W2}\sim25$, justifying the noticeably reduced number of sources compared to panel\,(b).

The two candidate BDs that survived the PM selection in Fig.\,\ref{PM} are marked as green triangles in panel\,(c), with error bars representing their colour uncertainty, as estimated above. These two sources fall in a region of the CMD where completeness is very low, below 10\% and approaching 0\%. The likelihood of accurately detecting sources in this magnitude range is, therefore, quite limited and, as a result, the identification of these objects as bona fide BDs is questionable. Given the extremely low completeness at these faint magnitudes, it is difficult to trust the reliability of these sources as genuine members of the cluster, making their classification as BDs rather uncertain.

\section{Comparison to model isochrones}\label{Section5}

\begin{centering} 
\begin{figure}
\includegraphics[width=\columnwidth]{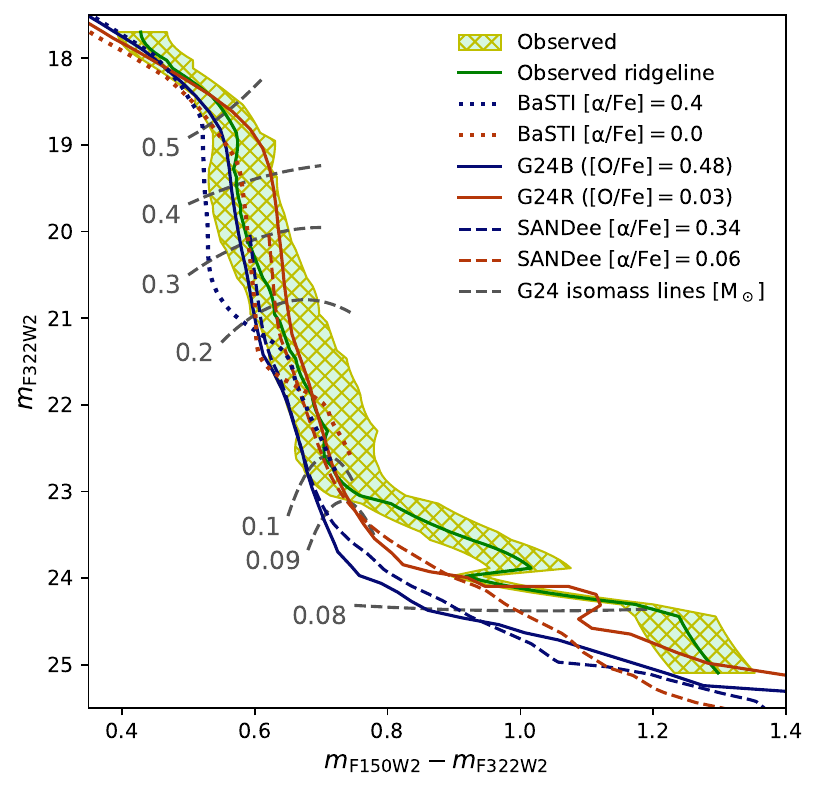}
\caption{Comparison of the observed colour distribution in 47\,Tuc with theoretical isochrones from three different sets of models. For each set, two isochrones are shown with distinct $[\mathrm{O/Fe}]$ to represent the mPOPs distribution of oxygen abundances in the cluster. The shaded area spans the 5th and the 95th percentiles of the colour spread at each magnitude with the median percentile (the ridgeline) shown in green. G24 refers to \citet{roman_47Tuc}.} 
\label{fig:isochrones} 
\end{figure} 
\end{centering}

\subsection{Model isochrones in literature}

Member-to-member variations in element abundances driven by the presence of mPOPs are responsible for the characteristically widened CMD in \cref{CMD,PM,comp,ASs,ASs2}. The $(m_{\rm F150W2}-m_{\rm F322W2})$ colour is particularly sensitive to the strength of infrared $\mathrm{H_2O}$ absorption and, hence, the oxygen abundance, $[\mathrm{O/Fe}]$. Capturing the observed colour distribution in the CMD requires at least two model isochrones with adopted values of $[\mathrm{O/Fe}]$ that represent the lower and upper bounds of the chemical spread in the cluster. In Fig.\,\ref{fig:isochrones}, we compare the observed $(m_{\rm F150W2}-m_{\rm F322W2})$ colour distribution to three sets of model isochrones in the literature.

\begin{centering} 
\begin{figure}
\includegraphics[width=\columnwidth]{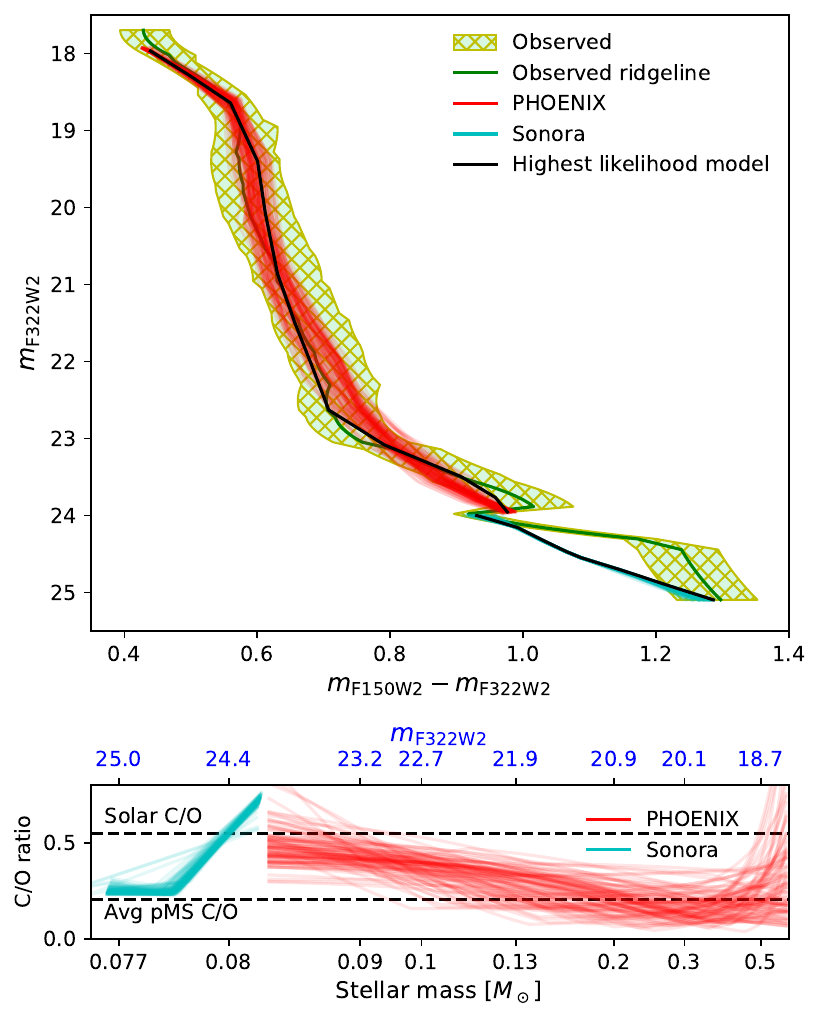}
\caption{\textit{Top:} CMD fits of the observed colour distribution in 47\,Tuc to models with abundances that vary with stellar mass. The red and cyan lines show the model CMDs from $100$ randomly chosen MCMC steps for \texttt{PHOENIX} and \texttt{Sonora} model atmospheres, respectively. In both cases, the step with the highest likelihood is shown in black. \textit{Bottom:} the $\mathrm{C/O}$ abundance ratio as a function of stellar mass for the same MCMC steps as in the top panel. The scaling of the mass axis is logarithmic in $(M-0.075\ M_\odot)$. The corresponding magnitudes are indicated in blue. The solar $\mathrm{C/O}$ ratio of $0.55$ and the average spectroscopically inferred 47\,Tuc $\mathrm{C/O}$ from post-MS (pMS) members of $0.20$ are shown with horizontal dashed lines, based on \citet{solar_CO} and \citet{nominal_C14,nominal_T14,nominal_M16}, respectively. The large values of $\mathrm{C/O}$ predicted by \texttt{PHOENIX} models at $M\gtrsim 0.5M_\odot$ are likely an artifact of the method, since our models only incorporate the effect of atmospheric chemistry that is expected to be subdominant at those large masses.}
\label{fig:variable_co} 
\end{figure} 
\end{centering}

Since the analysis carried out in this section is primarily concerned with the range of chemical abundances present in the cluster, and not with the specific shape of their distribution, we do not treat the CMD as a collection of individual stars. Instead, we consider the ridgeline and the tails of the overall colour distribution at each $m_{\rm F322W2}$ magnitude. To obtain those parameters, we divided the CMD into $20$ uniform magnitude bins. In each magnitude bin, we fitted a quadratic polynomial to the observed colour-magnitude trend and determined the required shifts in the best-fit polynomial along the colour axis that place $5\%$, $50\%$ and $95\%$ of stars in the magnitude bin on the blue side of the shifted polynomial. The three shifted polynomials stitched across all magnitude bins were then taken as the blue tail, the ridgeline, and the red tail of the colour distribution, respectively. In Figs.\,\ref{fig:isochrones} and \ref{fig:variable_co}, the part of the CMD contained between the two identified tails is shaded in light green, and the ridgeline is shown with the solid green line.

The model isochrones shown in Fig.\,\ref{fig:isochrones} are the following:

\begin{enumerate}
    \item \texttt{BaSTI} (a \texttt{B}ag of \texttt{S}tellar \texttt{T}racks and \texttt{I}sochrones, \citealt{BaSTI_1,BaSTI_2}) isochrones are available for $[\mathrm{O/Fe}]=[\mathrm{\alpha/Fe}]=0.0$ and $0.4$. These $\mathrm{O}$ abundances correspond to approximately the 25th and the 85th percentiles of the spectroscopically inferred distribution of $[\mathrm{O/Fe}]$ in 47\,Tuc \citep{nominal_C14,nominal_T14}. We, therefore, expect the metallicity-scaled solar and $\alpha$-enhanced \texttt{BaSTI} isochrones to capture a significant fraction of the colour spread in the CMD. We adopted the average spectroscopic metallicity of the cluster ($[\mathrm{Fe/H}]=-0.75$, \citealt{nominal_C14,nominal_T14,nominal_M16}). Unfortunately, \texttt{BaSTI} isochrones are only available for stellar masses $M\geq0.1\ M_\odot$, and therefore can only be compared to bright members with $m_\mathrm{F322W2}\lesssim 23$.
    \item In our previous work \citep{roman_47Tuc}, we presented a set of model isochrones that were specifically tailored to capture the chemical heterogeneity of 47\,Tuc. The overall metallicity, $[\mathrm{Fe/H}]$, and the abundances of $26$ elements were set to the average of the spectroscopic measurements in the literature. Two more elements, $\mathrm{O}$ and $\mathrm{Ti}$, were adjusted to obtain the best correspondence between the models and the optical and infrared HST photometry of the cluster. The set includes three isochrones that were intended to capture the blue tail, the red tail and the ridgeline of the CMD (hereafter referred to as \texttt{G24B}, \texttt{G24R} and \texttt{G24RL}, correspondingly). While the model isochrones extend down to $m_\mathrm{F322W2}\sim 28$, the calibration of chemical abundances was carried out on MS members with $M>0.1\ M_\odot$, corresponding to $m_\mathrm{F322W2}\lesssim 23$.
    \item \texttt{SANDee} (\texttt{e}volutionary \texttt{e}xtension to \texttt{SAND}, \citealt{2024ApJ...971...65G}) isochrones utilize the recently published grid of \texttt{SAND} (\texttt{S}pectral \texttt{AN}alog of \texttt{D}warfs, \citealt{efrain}) low-temperature model atmospheres to provide isochrones for a wide variety of $[\mathrm{Fe/H}]$ and $[\mathrm{\alpha/Fe}]$, spanning the range of effective temperatures from $T_\mathrm{eff}=700\ \mathrm{K}$ to $T_\mathrm{eff}=4000\ \mathrm{K}$. We compare our observations to the highest and lowest available $[\mathrm{\alpha/Fe}]$ within the convex hull of \texttt{SANDee} models at the average spectroscopic metallicity of the cluster ($[\mathrm{Fe/H}]=-0.75$, \citealt{nominal_C14,nominal_T14,nominal_M16}).
\end{enumerate}

For all model isochrones, we adopted the optical reddening of $E(B-V)=0.04$ and the age of $11.5\ \mathrm{Gyr}$ based on the isochrone fitting in \citet{roman_47Tuc}; the distance modulus of $13.24$ from \citet{dm}; and the spectroscopic metallicity of $[\mathrm{Fe/H}]=-0.75$. Reddening corrections were carried out using the \texttt{BasicATLAS} package \citep{mikaela} assuming the total-to-selective extinction ratio of $R_V=3.1$ and the reddening law from \citet{extinction}.

At the brightest magnitudes shown in Fig.\,\ref{fig:isochrones} ($m_\mathrm{F322W2}\lesssim 18.5$), all model isochrones are in excellent agreement with each other, yet they fail to capture the full width of the CMD. This indicates that at high stellar masses ($M\gtrsim0.5\ M_\odot$), the photometric scatter is primarily driven by effects other than atmospheric chemistry (e.g., interior helium mass fraction, multiple star systems).

The presence of mPOPs becomes apparent in the CMD at $m_\mathrm{F322W2}\gtrsim 19$ ($M\lesssim0.5\ M_\odot$), as both the observed MS and the range of colours predicted by the model isochrones rapidly widen and atmospheric chemistry becomes the dominant contributor to the observed spread in photometry. At intermediate masses ($0.3\ M_\odot\lesssim M\lesssim 0.5\ M_\odot$), the cumulative photometric scatter predicted by \texttt{BaSTI} isochrones and \citet{roman_47Tuc} is in agreement with observations; however, neither of the two sets of isochrones is able to reproduce the scatter on its own: \texttt{BaSTI} isochrones appear biased toward bluer colours, while \citet{roman_47Tuc} isochrones primarily capture the red tail of the colour distribution. This is not surprising, as the chemical abundances adopted in \texttt{BaSTI} are $[\mathrm{Fe/H}]$/$[\mathrm{\alpha/Fe}]$-scaled solar and do not capture the chemical intricacies of 47\,Tuc, while the isochrones in \citet{roman_47Tuc} were calibrated at shorter wavelengths where the spectral energy distribution may not be sufficiently sensitive to the abundances of elements that impact the infrared CMD.

Below $0.3\ M_\odot$ ($m_\mathrm{F322W2}\gtrsim 20.5$), no set of model isochrones captures the red tail of the distribution. The disagreement becomes more prominent at fainter magnitudes, resulting in a complete departure of the models from the observed colours near $M=0.1\ M_\odot$. At $m_\mathrm{F322W2}\sim 24$, the observed MS appears to display a discontinuity, as it suddenly narrows, ``steps down'' towards fainter magnitudes and bluer colours, and changes its slope. This discontinuity, first observed by \citet{2024ApJ...965..189M}, is not matched by any of the model isochrones, although the \texttt{G24R} isochrone from \citet{roman_47Tuc} appears to display a similar ``step down'' some $0.5$ magnitudes fainter at a redder colour. In Sec. \ref{sec:discontinuity}, we argue that the observed discontinuity and the ``step down'' in the \texttt{G24R} isochrone have the same physical origin, but are offset from one another in the CMD due to the overestimated oxygen abundance in the model and incomplete opacity treatment.

\subsection{Variations of abundances with stellar mass}

The disagreement between theoretical isochrones and observations in Fig.~\ref{fig:isochrones} is qualitatively similar to that discovered by \citet{mattia_paper} in the globular clusters NGC 6121 and NGC 6397. \citet{mattia_paper} found that in all combinations of the JWST Near Infrared Imager and Slitless Spectrograph (NIRISS) F150W band and HST Advanced Camera for Surveys (ACS) F606W and F814W bands, the models consistently predict bluer colours than observed at fainter magnitudes near the end of the MS. The authors examined a number of possible systematic errors including uncertain convective parameters and rates of dust formation. They found all considered effects to be too small to cause the observed discrepancy.

We note that \citet{mattia_paper} did not consider potential shortcomings in the chemical equilibrium and opacity treatment adopted in their model atmospheres. It is unfortunately not uncommon for different sets of models to predict vastly different spectral energy distributions for stars with $T_\mathrm{eff}\ll3000\ \mathrm{K}$. The issue is often more pronounced at sub-solar metallicities, as the calibration sample of spectroscopically observed metal-poor cool stars remains small \citep{subdwarf_hunt_1,subdwarf_hunt_2}. For instance, \citet{speedy_Ldwarf} recently found a discrepancy of $\sim 500\ \mathrm{K}$ and $\sim0.5\ \mathrm{dex}$ in the best-fit $T_\mathrm{eff}$ and $[\mathrm{Fe/H}]$, respectively, when comparing the observed spectrum of a nearby L subdwarf to different grids of model atmospheres. It is nonetheless important to emphasize that the isochrones shown in Fig.~\ref{fig:isochrones} fail to capture the red tail of the colour distribution in the CMD at stellar masses as high as $0.28\ M_\odot$, corresponding to much warmer temperatures ($T_\mathrm{eff}\gg 3000\ \mathrm{K}$), at which model atmospheres are expected to be far less sensitive to the adopted approximations.

In this section, we attempt to attain a better agreement between the observed CMD of 47\,Tuc and theoretical isochrones by adjusting the input parameters of model atmospheres. That is, we assume that the models themselves are free of major systematic errors. Without major alterations in the physical simulations, it is challenging to reproduce the observed colour spread without allowing some variation in the chemical abundances with stellar mass. Such variation would be at odds with previous studies that established tight constraints on the mass dependence of chemical composition \citep{no_mass_variations_1,2023MNRAS.522.2429M,no_mass_variations_3}, including our earlier analysis of 47\,Tuc \citep{roman_47Tuc}. We attribute this contradiction to three effects. First, previous analyses had limited depth, at which the discrepancy between the observed colour spread and model isochrones is less significant. For example, both the $(m_\mathrm{F115W}-m_\mathrm{F322W2})$ and $(m_\mathrm{F606W}-m_\mathrm{F115W})$ colours presented in \citep{2023MNRAS.522.2429M} seem to display a rapid reddening of the MS at $m_\mathrm{F115W}\gtrsim 23$ similar to the one in Fig.~\ref{fig:isochrones}; however, this effect falls beyond the faint limit of the theoretical isochrones considered in that study. Second, the discrepancy appears to be far less prominent in certain colour combinations such as HST Wide Field Camera 3 (WFC3) $(m_\mathrm{F110W}-m_\mathrm{F160W})$ examined in \citet{no_mass_variations_3} and \citet{roman_47Tuc}. Third, while the discrepancy is readily seen in the optical $(m_\mathrm{F606W}-m_\mathrm{F814W})$ colour as demonstrated in \citet{mattia_paper}, this colour is also sensitive to $\mathrm{TiO}$ absorption bands and, in turn, the adopted value of $[\mathrm{Ti/Fe}]$. Isochrones with large $\mathrm{Ti}$ or $\alpha$-enhancement may therefore conceal the discrepancy in this particular colour combination. We suspect that the large value of $[\mathrm{Ti/Fe}]=0.64$ measured in 47\,Tuc by \citet{roman_47Tuc} (compared to the spectroscopic measurement of $[\mathrm{Ti/Fe}]=0.3$ from \citealt{nominal_C14,nominal_T14}), and the large value of $[\mathrm{\alpha/Fe}]=0.6$ measured in $\omega$\,Cen by \citet{roman_omega_cen} (compared to the spectroscopic measurement of $[\mathrm{Ti/Fe}]=0.2$ from \citealt{omega_cen_spectroscopic_abundances}) are symptomatic of this issue.

To explore possible variations of chemical abundances as a function of stellar mass in 47\,Tuc, we calculated a new grid of model atmospheres using \texttt{PHOENIX 15} \citep{phoenix_origin,BT-Settl,phoenix_15,roman_note}. The grid comprises four dimensions of parameters: stellar mass (between $0.082$ and $0.6\ M_\odot$), $[\mathrm{O/Fe}]$ (between $-0.2$ and $0.7\ \mathrm{dex}$), $[\mathrm{C/Fe}]$ (between $-0.5$ and $0.25\ \mathrm{dex}$) and $[\mathrm{Ti/Fe}]$ (between $0$ and $0.64\ \mathrm{dex}$). The choice of $\mathrm{O}$ and $\mathrm{Ti}$ is motivated by their large effect on the Rosseland mean opacity of the atmosphere as detailed in \citet{roman_47Tuc}. $[\mathrm{C/Fe}]$ is included to capture the effects of carbon-dominated chemistry, enabled by the suppressed formation of $\mathrm{CO}$ in oxygen-poor atmospheres \citep{CO_ratio}. All other abundances were set to the values adopted in \citet{roman_47Tuc}. The stellar masses were converted to $T_\mathrm{eff}$ and $\log(g)$ using the isochrones from \citet{roman_47Tuc}. Namely, we used the \texttt{G24RL} isochrone for models within the range of $[\mathrm{O/Fe}]$ considered in \citet{roman_47Tuc} ($0.03\leq [\mathrm{O/Fe}]\leq 0.48$), the \texttt{G24B} isochrone for models with higher oxygen abundances, and the \texttt{G24R} isochrone for models with lower oxygen abundances.

At stellar masses below $0.082$, we used the published grid of \texttt{Sonora Elf Owl} model atmospheres \citep{ElfOwl}, bicubically interpolated to $T_\mathrm{eff}$ and $\log(g)$ predicted by the \texttt{G24RL} isochrone from \citet{roman_47Tuc}. The grid spans three dimensions (other than $T_\mathrm{eff}$ and $\log(g)$): $[\mathrm{M/H}]=[\mathrm{Fe/H}]$ (between $-1$ and $+1\ \mathrm{dex}$), $\mathrm{C/O}$ abundance ratio (between $0.22$ and $1.14$) and the vertical diffusion parameter, $\log(K_{zz})$ (from $2$ to $9$ in cgs units). $\log(K_{zz})$ regulates non-equilibrium chemistry that may have a significant impact on the colours of cool stars due to the redistribution of key absorbers in the atmosphere.

The synthetic spectra of both \texttt{PHOENIX} and \texttt{Sonora} atmospheres were converted into model CMDs using the stellar radii predicted by the isochrones in \citet{roman_47Tuc}. We then fitted the model CMDs to the observed colour distribution of 47\,Tuc using the Markov Chain Monte Carlo (MCMC) approach. We allowed $[\mathrm{O/Fe}]$ and $[\mathrm{C/Fe}]$ in \texttt{PHOENIX} models, and $\mathrm{C/O}$ and $\log(K_{ZZ})$ in \texttt{Sonora} models to vary linearly with stellar mass. Each of the four parameters was allowed to have one change of slope at some best-fit intermediate mass. On the other hand, $[\mathrm{Ti/Fe}]$ in \texttt{PHOENIX} and $[\mathrm{Fe/H}]$ in \texttt{Sonora} were required to have no mass dependence.

The MCMC sampling was carried out with $32$ walkers in $10,000$ steps per walker using the Goodman-Weare algorithm \citep{MCMC} implemented in \texttt{emcee}. The sampling for \texttt{PHOENIX} and \texttt{Sonora} models was done separately. At each step, the log-likelihood was evaluated assuming that the $(m_{\rm F150W2}-m_{\rm F322W2})$ colour in 47\,Tuc follows an asymmetric normal distribution at each $m_{\rm F322W2}$ magnitude, with the median of the distribution placed at the 50th percentile in the observed scatter in colour at that magnitude, and the left and right two-sigma tails of the distribution placed at the 5th and the 95th percentiles of the observed scatter. The upper panel of Fig\,\ref{fig:variable_co} shows the model CMDs from $100$ steps, randomly chosen from the second half of the calculated MCMC chains. The panel also shows the CMDs from the steps with the highest evaluated log-likelihood. The lower panel of the figure shows the variation of the $\mathrm{C/O}$ ratio as a function of stellar mass for the same $100$ steps. The ratio was inferred directly from the \texttt{Sonora} models, and calculated from $[\mathrm{O/Fe}]$ and $[\mathrm{C/Fe}]$ for the \texttt{PHOENIX} models.

Table\,\ref{table:MCMC_params} lists the ranges of values of the adopted model parameters compatible with the observed CMD at the lowest and highest masses within the fitting range of each model set. The values shown in the table were inferred from the 5th and the 95th percentiles of the marginalized MCMC posteriors. In cases where one of the tails of the posterior distribution overflowed the bounds of the model grid, the range is shown as either the lower 5th percentile bound or the upper 95th percentile bound. The posterior distributions where both tails overflowed the bounds are listed as ``unconstrained''.

At the highest mass in the fitting range ($0.6\ M_\odot$), the inferred upper bound on the distribution of $[\mathrm{O/Fe}]$ is in good agreement with the spectroscopic measurements \citep{nominal_C14,nominal_T14}. Note that the lower bound is not constrained, since the red tail of the CMD at $M=0.6\ M_\odot$ is likely driven by effects other than atmospheric chemistry, and therefore it cannot be reproduced with our model atmospheres. The carbon abundance at $0.6\ M_\odot$ is completely unconstrained, as most carbon atoms are locked in the $\mathrm{CO}$ molecule in oxygen-rich atmospheres, and therefore the impact of $[\mathrm{C/Fe}]$ on the observed colour is minimal.

For \texttt{PHOENIX} atmospheres at $M\sim0.082$, all values of $[\mathrm{O/Fe}]$ higher than $-0.1$ are incompatible with the observed colour, suggesting that the majority of cluster members near the end of the MS are oxygen-poor. The increased sensitivity of oxygen-poor atmospheres to $[\mathrm{C/Fe}]$ also allows us to derive constraints on the carbon abundance. Notably, the inferred upper limit on $[\mathrm{C/Fe}]$ of $-0.02\ \mathrm{dex}$ suggests that the $\mathrm{C/O}$ ratio remains below $1$ even at the lowest oxygen abundance of $-0.2$ allowed by the model grid (for the solar abundances adopted in our \texttt{PHOENIX} models, $\mathrm{C/O}=1$ occurs when $[\mathrm{C/Fe}]-[\mathrm{O/Fe}]=0.26$). This behavior is expected, as over-production of carbon-bearing molecules (most importantly, $\mathrm{CH_4}$) in carbon-dominated atmospheres would suppress the flux in the \texttt{F322W2} band, resulting in a much bluer colour than observed.

The \texttt{Sonora} models at $M\sim0.081$ yield a similar constraint on the $\mathrm{C/O}$ ratio, requiring it to be distinctly higher than the post-MS spectroscopic counterpart ($\mathrm{C/O}\sim0.2$, \citealt{nominal_C14,nominal_T14,nominal_M16}), but well below unity. At the lowest mass within the observed range ($\sim0.077\ M_\odot$), the \texttt{Sonora} models instead require a much lower $\mathrm{C/O}$ ratio comparable to the spectroscopic estimate. The low value of $\mathrm{C/O}$ is required to suppress the production of $\mathrm{CH_4}$ at lower $T_\mathrm{eff}$ in order to capture the observed colour distribution near $m_\mathrm{F322W2}=25$. However, the reliability of this estimate of $\mathrm{C/O}$ is questionable, as the \texttt{Sonora} CMD fails to capture the observed colour spread between $m_\mathrm{F322W2}\approx24$ and $m_\mathrm{F322W2}\approx25$, and requires the maximum allowed $\log(K_{zz})=9$ at all stellar masses. We also note that if the shift towards lower oxygen abundances at lower stellar masses is a real effect, the mass-radius and mass-temperature relationships adopted from the theoretical isochrones in \citet{roman_47Tuc} may be inappropriate at low $T_\mathrm{eff}$.

\begin{table}
\caption{Ranges of model parameters compatible with the observed CMD of 47\,Tuc.}
\label{table:MCMC_params}
\centering
\begin{tabular}{c l c c}
\hline
Model set & Parameter & Range $[\mathrm{dex}]$\\
\hline
   \texttt{PHOENIX} & $[\mathrm{Ti/Fe}]$ at all $M$ & Unconstrained \\
   \texttt{PHOENIX} & $[\mathrm{C/Fe}]$ at $0.600\ M_\odot$ & Unconstrained \\
   \texttt{PHOENIX} & $[\mathrm{O/Fe}]$ at $0.600\ M_\odot$ & $<0.50$ \\
   \texttt{PHOENIX} & $[\mathrm{C/Fe}]$ at $0.082\ M_\odot$ & $-0.36 \rightarrow -0.02$ \\
   \texttt{PHOENIX} & $[\mathrm{O/Fe}]$ at $0.082\ M_\odot$ & $<-0.10$ \\
   \texttt{Sonora} & $[\mathrm{Fe/H}]$ at all $M$ & $-0.50 \rightarrow -0.48$ \\
   \texttt{Sonora} & $\mathrm{C/O}$ at $0.081\ M_\odot$ & $0.55 \rightarrow 0.69$ \\
   \texttt{Sonora} & $\log(K_{zz})$ at $0.081\ M_\odot$ & $8.84 \rightarrow 8.99$ \\
   \texttt{Sonora} & $\mathrm{C/O}$ at $0.077\ M_\odot$ & $0.15 \rightarrow 0.24$ \\
   \texttt{Sonora} & $\log(K_{zz})$ at $0.077\ M_\odot$ & $>8.36$ \\
\hline
\end{tabular}
\end{table}

\subsection{Main sequence discontinuity: the low-mass MS ``kink''}
\label{sec:discontinuity}

In this section, we address the apparent discontinuity in the MS of 47\,Tuc observed near $m_\mathrm{F322W2}=24$. We will refer to this feature as to the \textit{``kink''} of the low-mass MS. The \textit{kink} is unlikely to be an artefact of our data reduction, since no equivalent discontinuity is observed in the MS of the SMC (see Fig.\,\ref{CMD}), and since this feature has also been reported in \citet{2024ApJ...965..189M} (see the upper teal arrow in their Fig.\,5).

The \texttt{G24R} isochrone from \citet{roman_47Tuc}, plotted in Fig\,\ref{fig:isochrones}, appears to display a similar discontinuity in the slope, albeit at a fainter magnitude ($m_\mathrm{F322W2}=24.5$ as opposed to $24$) and redder colour ($m_\mathrm{F150W2}-m_\mathrm{F322W2}=1.1$ as opposed to $0.9$). The discontinuity in the isochrone is caused by the rapid onset of $\mathrm{CH_4}$ absorption in the \texttt{F322W2} band at $T_\mathrm{eff}\lesssim2000\ \mathrm{K}$ that, for a narrow range of temperatures between $1900\ \mathrm{K}$ and $2000\ \mathrm{K}$, inverts the colour-$T_\mathrm{eff}$ relationship, making cooler stars appear bluer than their warmer counterparts. This transition is driven by the depletion of oxygen atoms onto dust grains that, in turn, suppresses the formation of $\mathrm{CO}$ and releases additional carbon atoms into the atmosphere to produce $\mathrm{CH_4}$. After the transition, the atmosphere is less sensitive to the oxygen abundance, which explains the narrow scatter in colour below the discontinuity.

The \texttt{G24B} isochrone, also shown in Fig\,\ref{fig:isochrones}, does not display a similar feature due to its much higher $[\mathrm{O/Fe}]$. As a result, the onset of $\mathrm{CH_4}$ absorption happens at a lower $T_\mathrm{eff}$ ($\sim1700\ \mathrm{K}$). In general, the conversion of $\mathrm{CO}$ to $\mathrm{CH_4}$ is favoured at lower temperatures, making the transition more gradual and free of discontinuities. Since the discontinuity occurs over a very narrow range of $T_\mathrm{eff}$, it can be overlooked in model grids with sparse temperature sampling. The isochrones in \citet{roman_47Tuc} were calculated using the so-called hammering method (see their Section~4.7) with model atmospheres calculated in steps of $\Delta T_\mathrm{eff}=50\ \mathrm{K}$ in key regions of the parameter space. On the other hand, \texttt{SANDee} isochrones (also shown in Fig.\,\ref{fig:isochrones}) were obtained using the standard method, where a sparse grid of model atmospheres is linearly interpolated to the $T_\mathrm{eff}$ and $\log(g)$ predicted by the evolutionary models. This is the most likely reason why the $[\mathrm{O/Fe}]=0.06$ \texttt{SANDee} isochrone does not display the discontinuity despite its similar chemical composition to the \texttt{G24R} isochrone.

If the rapid onset of $\mathrm{CH_4}$ absorption driven by the low oxygen content of the atmosphere is indeed the cause of the observed discontinuity, then why does the model place this feature at a fainter magnitude and redder colour? Table~\ref{table:MCMC_params} suggests that near the \textit{kink}, $\mathrm{[O/Fe]}$ is likely much lower than that adopted in the \texttt{G24R} isochrone. In an atmosphere with $\mathrm{[O/Fe]}<0$, the onset of $\mathrm{CH_4}$ absorption is expected to occur at $T_\mathrm{eff}>2000\ \mathrm{K}$, corresponding to a bluer colour and brighter magnitude, in agreement with the observed CMD. This hypothesis may be tested in a future study that reproduces the modelling approach adopted in \citet{roman_47Tuc} for a set of oxygen abundances lower than $\mathrm{[O/Fe]}=0.0$. We also note that this offset may be due to a poorly modelled source of continuum opacity in \texttt{G24R} that would make the absolute colours and magnitudes predicted by the isochrone less accurate. There is however no doubt that the observed \textit{kink} can be reproduced by the current generation of model isochrones only if the cluster members near the end of the main sequence are disproportionately more oxygen-poor than their higher-mass counterparts on the upper main sequence or the red giant branch.

\begin{centering} 
\begin{figure}
\includegraphics[width=\columnwidth]{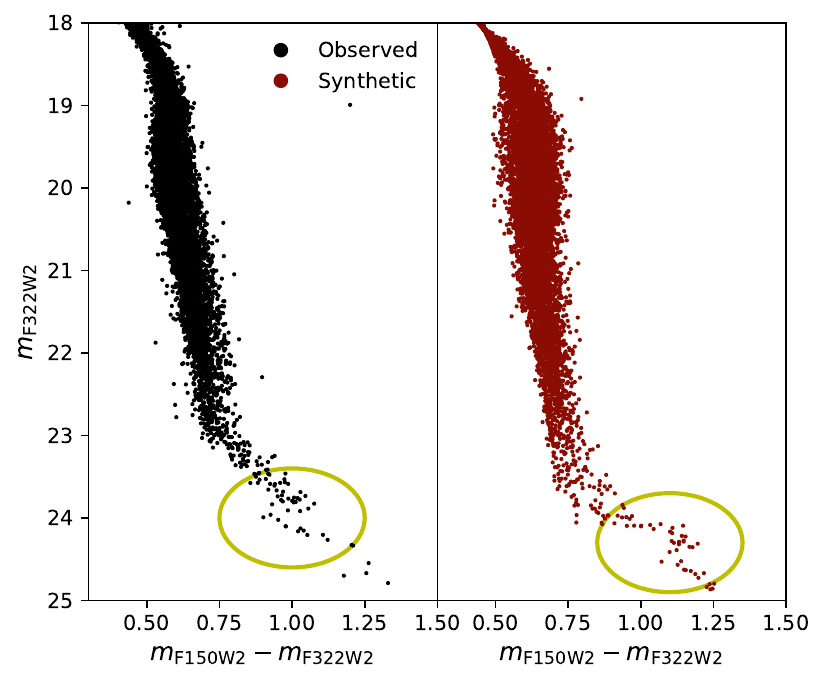}
\caption{Comparison of the membership-selected observed CMD of 47\,Tuc (\textit{left}), and the equivalent synthetic CMD, produced with the theoretical isochrones from \citet{roman_47Tuc} as described in text. Both CMDs display a similar discontinuity highlighted with yellow ellipses.} 
\label{simulated_discontinuity} 
\end{figure} 
\end{centering}

The emergence of the discontinuity is demonstrated in Fig.~\,\ref{simulated_discontinuity}. The left panel of the figure shows the observed CMD of 47\,Tuc. The right panel shows a synthetic CMD with identical plot limits. To produce the latter, we first drew $12,000$ random initial masses between $0.07\ M_\odot$ and $0.6\ M_\odot$ from the broken power law distribution, derived for 47\,Tuc in \citet{roman_47Tuc}. Next, we drew a random oxygen abundance ($[\mathrm{O/Fe}]$) for each initial mass from the normal distribution with $(\mu=0.2,\sigma=0.3)$ at $0.085<M/M_\odot<0.6$, $(\mu=-0.2,\sigma=0.05)$ at $0.081\ M_\odot$, and linearly interpolated $\mu$ and $\sigma$ at other initial masses. The generated masses and abundances were converted into colours and magnitudes using linearly interpolated and extrapolated isochrones \texttt{G24R}, \texttt{G24RL} and \texttt{G24B}. Finally, a fraction of the synthetic sample was removed at random according to the magnitude-dependent completeness derived in Sec.~\ref{Section6}. The completeness relationship was shifted by $0.5\ \mathrm{mag}$ towards fainter magnitudes to account for the systematic offset between the observed magnitude of the discontinuity, and the predicted magnitude of $\mathrm{CH_4}$ absorption in \texttt{G24R}. The completeness correction reduced the synthetic sample to approximately the same number of stars as that in the observed dataset ($\sim9000$). The final synthetic CMD in the right panel of Fig.~\,\ref{simulated_discontinuity} is visually similar to the observed \textit{kink}, despite appearing at a redder colour and fainter magnitude.

\section{Luminosity function}\label{Section6}

\begin{centering} 
\begin{figure}
\includegraphics[width=\columnwidth]{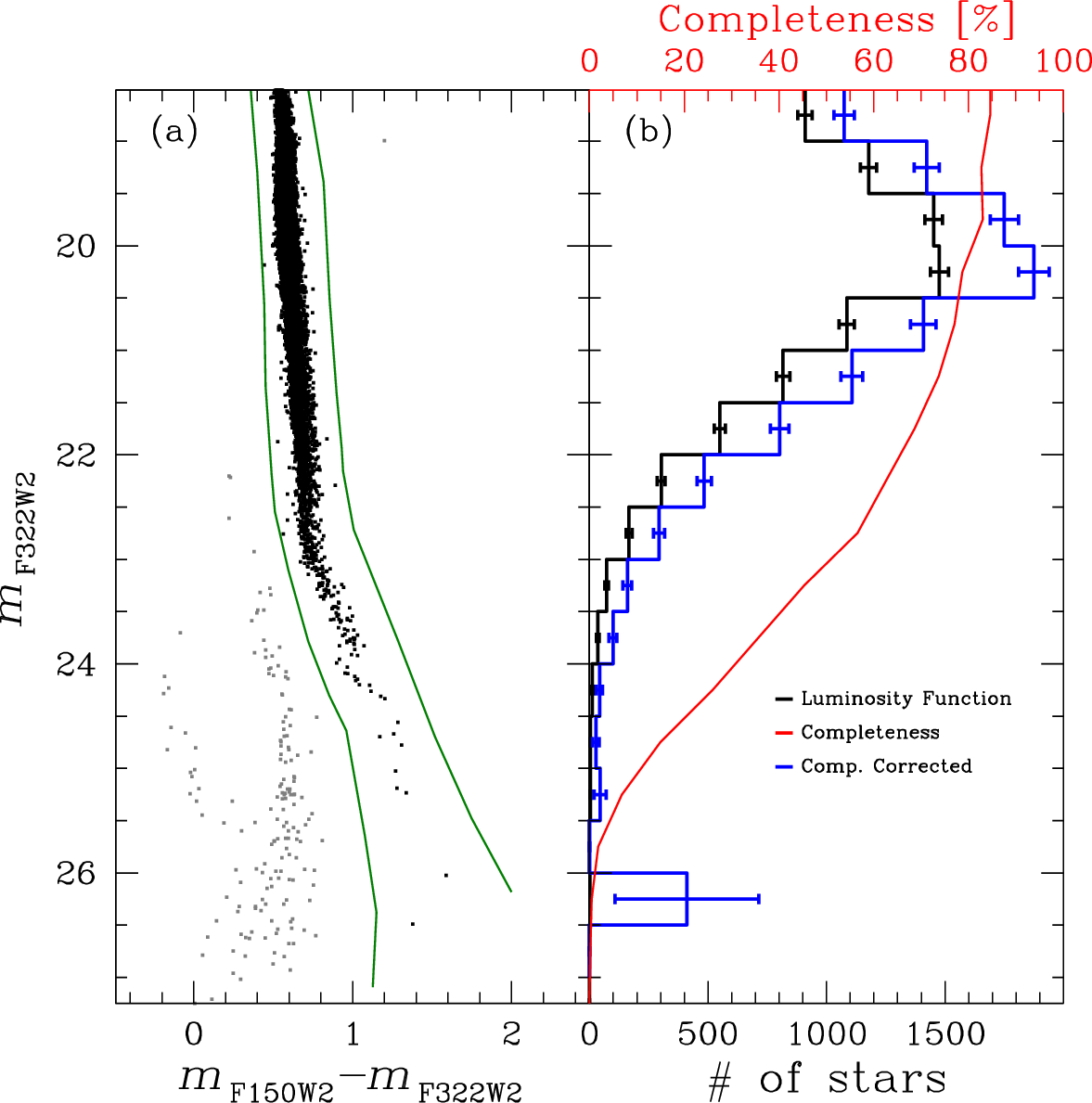}
\caption{Luminosity function of 47\,Tuc. (a) $m_{\rm F322W2}$ versus $m_{\rm F150W2}-m_{\rm F322W2}$ CMD for our selected sample of stars. The two green lines are used to isolate the 47\,Tuc MS stars. Sources between the two green lines are represented in black, while other stars are represented in grey. (b) Number counts of stars (in black), binned into 0.5 $m_{\rm F322W2}$ magnitude intervals, corresponding to the black points in (a). The red line shows the completeness as a function of magnitude, with values reported at the top of the plot. The completeness-corrected LF is represented in blue.} 
\label{LF} 
\end{figure} 
\end{centering}

In this section, we present the LF of 47\,Tuc in the lower part of the MS. Panel\,(a) of Fig.\,\ref{LF} shows the $m_{\rm F322W2}$ versus $m_{\rm F150W2}-m_{\rm F322W2}$ CMD for our selected sample of stars. We made use of the two fiducial lines defined in panel\,(a) of Fig.\,\ref{ASs} to isolate the MS of 47\,Tuc (represented with black dots) from other stars (represented with grey dots). The number count of selected stars, binned in 0.5\,$m_{\rm F322W2}$ magnitude intervals, is presented as a black histogram in panel\,(b), with the error bars indicating Poisson uncertainties. The red line in the same panel illustrates the completeness as a function of magnitude, with the corresponding value displayed at the top. The completeness-corrected LF is represented in blue, with errors estimated by propagating the uncertainties.

A noteworthy feature of the LF is the presence of a bump at the magnitude bin corresponding to the two BD candidates surviving the PM selection. However, as discussed in earlier sections, the significant uncertainties in the PM measurements at these faint magnitudes, coupled with the very low completeness of the data in this regime, make it unlikely that these two sources are reliable cluster members or bona fide BDs. Given these uncertainties, we cannot confidently attribute this bump to an actual population of BDs within the cluster.

\begin{centering} 
\begin{figure}
\includegraphics[width=\columnwidth]{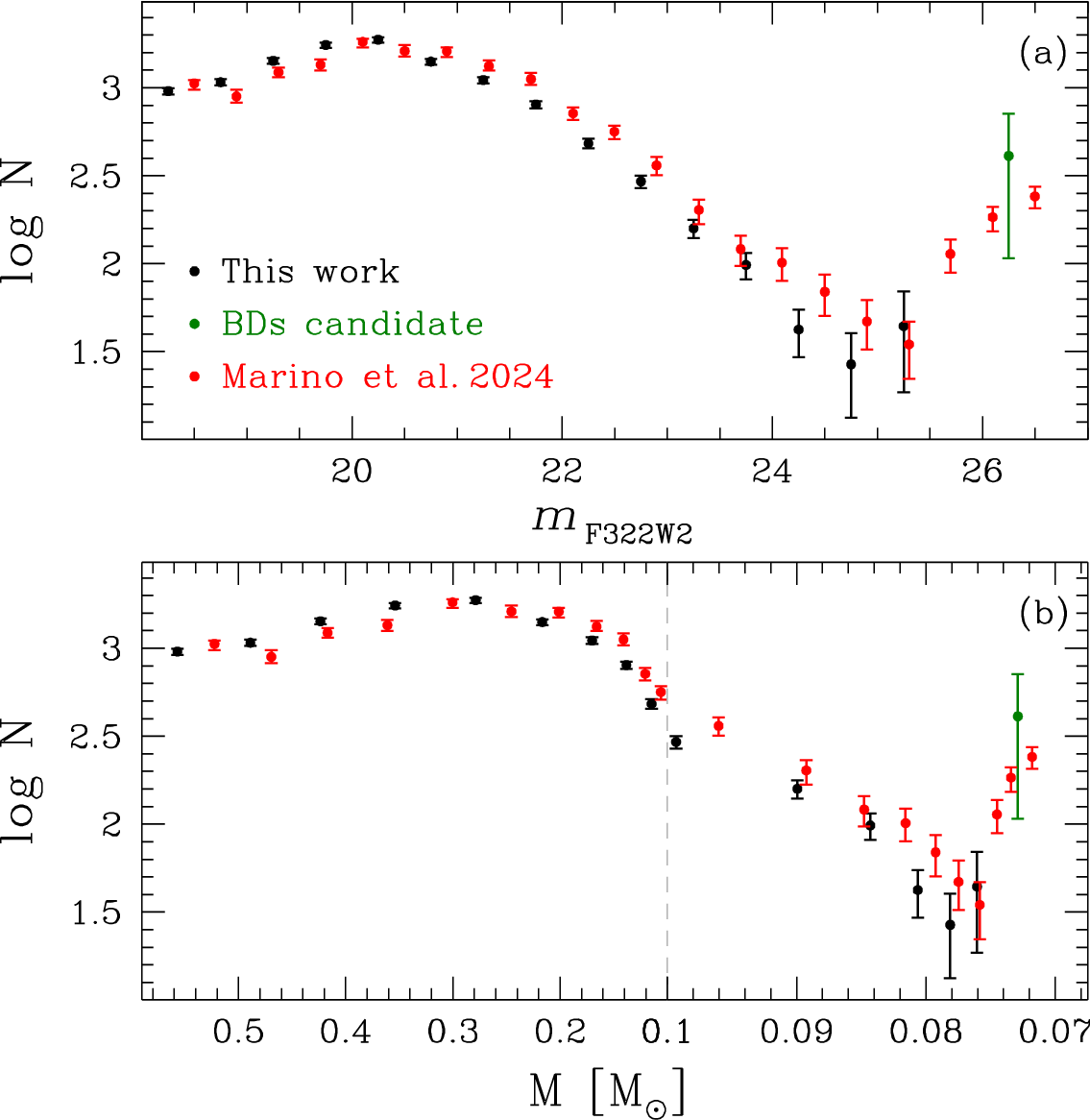}
\caption{Comparison between the LF (panel a) and MF (panel b) obtained in this work (black) with those presented in \citet{2024ApJ...965..189M} (red). The bin associated with the two suspected BDs in our work is highlighted in green in both panels.} 
\label{LF_MF} 
\end{figure} 
\end{centering}

In panel\,(a) of Fig.\,\ref{LF_MF}, we compare the LF derived in this work with the LF from \citet{2024ApJ...965..189M}, normalized to align with our LF. We used the isochrone from Fig.\,\ref{fig:isochrones} to convert magnitudes to solar masses for both LFs. Panel\,(b) shows the derived MFs. The bin associated with the two suspected BDs is highlighted in green in both panels.

Both LFs exhibit a similar trend, peaking around $m_{\rm F332W2} \sim 20.25$ and then gradually decreasing for fainter magnitudes. The decline in the LF from \citet{2024ApJ...965..189M} is less steep, with values slightly higher than in our LF. The minimum in our LF occurs earlier, at $m_{\rm F332W2} \sim 24.75$, compared to $m_{\rm F332W2} \sim 25.3$ in the LF from \citet{2024ApJ...965..189M}. After reaching this minimum, our LF shows a slight increase around $m_{\rm F332W2} \sim 25.25$ before dropping to zero, excluding the green bin, representing the two BD candidates. As discussed earlier, due to large proper-motion errors and low completeness at these faint magnitudes, these sources likely represent field stars rather than definitive BDs. In contrast, the LF from \citet{2024ApJ...965..189M} continues to rise beyond its minimum at $m_{\rm F332W2} \sim 25.3$. This may be due to the unavailability of PMs in their study, which likely resulted in the inclusion of background galaxies and field stars in their LF.

\section{Summary}\label{Section7}

In this work, we have conducted a detailed analysis of the GC 47\,Tuc using JWST data, providing new insights into the cluster’s low-mass stars and brown dwarfs. Our key findings are summarized as follows:

\begin{enumerate}

\item We present, for the first time, a proper-motion decontamination using two JWST epochs separated by just $\sim$7 months. This method proves to be effective down to $m_{\rm F322W2} \sim 27$ for the available data set of 47\,Tuc. New observations planned for 2025 under JWST proposal GO-5896 \citep{2024jwst.prop.5896S} will extend the time baseline by a factor of $\sim$ 4, providing approximately 3 years of data. This extended baseline will allow us to explore the hydrogen-burning limit well into the BD sequence.

\item We do not find strong or conclusive evidence for an increase in the LF associated with the presence of BDs at faint magnitudes. The unavailability of proper-motion decontamination in \citet{2024ApJ...965..189M} may have contributed to the inclusion of faint, red background galaxies in their BD LF. Our analysis aims to address this concern and provides a refined representation of the LF. A more detailed characterization will be achieved once the additional GO-5896 observations are available.

\item We find that the CMD of 47\,Tuc near the end of the MS cannot be reproduced by theoretical isochrones unless the oxygen abundance in stellar atmospheres is allowed to vary with mass. Preliminary analysis using two different sets of models suggests that for stars with $M \sim 0.08 M_\odot$, the upper limit of $[\mathrm{O/Fe}]$ is about 0.6 dex lower than the value inferred for more massive stars ($M > 0.1 M_\odot$), both spectroscopically and photometrically.

\item We report a discontinuity in the low-mass MS of 47\,Tuc around $m_{\rm F322W2} = 24$, already reported by \citet{2024ApJ...965..189M}, characterized by an abrupt "step down" to fainter magnitudes and bluer colours. We associate this feature, named 'kink', with a theoretical prediction in oxygen-poor atmospheres, where the sudden onset of $\mathrm{CH_4}$ absorption occurs as more carbon atoms are released due to the suppressed formation of carbon monoxide. This discontinuity serves as an independent confirmation of reduced oxygen abundance in low-mass stars (below $0.1 M_\odot$) in the cluster.

\end{enumerate}

With the arrival of new data from the GO-5896 JWST programme, we expect to further refine these findings, particularly in the interest of BD characterization and the study of low-mass stars. These observations will enhance our ability to perform precise proper-motion measurements and provide deeper insight into the atmospheric and evolutionary properties of sub-stellar objects and the faintest stars in this GC.\\~ 

\begin{acknowledgements}
We dedicate this paper to the memory of our colleague Prof.\,Harvey Richer ($\star$ April 1944 ---$\dagger$ 13 November 2023), a highly accomplished astronomer and expert in stellar populations and in particular within globular clusters, who passed away during this project. Harvey grew up in Montreal and was at least the second star man to graduate from his high school, having been preceded by William Shatner by more than a decade. He worked at the University of British Columbia for most of his career, and his focus was the late stages of stellar evolution, in particular carbon stars and white dwarfs. We thank the referee for his valuable suggestions and comments, which helped improve the paper, as well as for his prompt revision.
\end{acknowledgements}

\bibliographystyle{aa}
\bibliography{main.bib}

\end{document}